\documentclass[twocolumn]{aastex62}
\usepackage{amsmath}
\usepackage{graphicx}
\usepackage{textcomp}
\usepackage{gensymb}
\usepackage{natbib}
\usepackage{longtable}
\usepackage{multirow}
\usepackage[caption=false]{subfig}
\usepackage{float}
\usepackage{enumitem}
\usepackage{rotating}

\begin{document}
\def\teff{T$_{eff}$}

\graphicspath{{figures/}}

\received{13 January 2021}
\revised{22 February 2021}
\accepted{17 March 2021}
\submitjournal{ApJ}

\shorttitle{Undetected Binaries Cause Age Gradients}
\shortauthors{Sullivan \& Kraus}

\title{Undetected Binaries Cause an Observed Mass Dependent Age Gradient in Upper Scorpius}

\author[0000-0001-6873-8501]{Kendall Sullivan}
\altaffiliation{NSF Graduate Research Fellow}
\affil{University of Texas at Austin, Austin TX 78712, USA}

\author{Adam L. Kraus}
\affil{University of Texas at Austin, Austin TX 78712, USA}

\correspondingauthor{Kendall Sullivan}
\email{kendallsullivan@utexas.edu}

\begin{abstract}
Young stellar associations represent a key site for the study of star formation, but to accurately compare observations to models of stellar evolution, the age of an association must be determined. The Upper Scorpius region is the youngest section of the Scorpius-Centaurus OB association, which is the largest collection of nearby, young, low-mass stars. The true age of Upper Scorpius is not clear, and an observed mass-dependent age gradient in Upper Scorpius, as well as in other star-forming regions, complicates age measurements. The age gradient may indicate a genuine astrophysical feature or may be an artifact of unrecognized systematic effects in stellar age measurements. We have conducted a synthetic red-optical low-resolution spectroscopic survey of a simulated analog to the Upper Scorpius star-forming region to investigate the effects of unresolved binary stars (which have mass-dependent demographics) on age measurements of a stellar population. We find that the observed mass-dependent age gradient in Upper Scorpius can be explained by a population of undetected binary stars. For a simulated population with an age of 10 (RMS = 2) Myr, we measure an age of 10.5 (RMS = 3.5) Myr for F stars, and 7.5 (RMS = 5.8) Myr for M stars. This discrepancy is caused by the mass-dependent mass ratio distribution and the variable steepness of the mass-luminosity relation. Our results support the previously suggested 10 Myr age for Upper Scorpius, with a small intrinsic age spread.
\end{abstract}

\section{Introduction} \label{sec:intro}
Most young stars exist in loosely- or unbound associations \citep{Miller1978, Lada2003, Bressert2010}, the products of star formation within a since-dissipated molecular cloud. These large populations of nominally coeval young stars are useful testbeds for studying the outcomes of star formation. A key attribute of stellar associations is their age, which is used to connect observations with theoretical models of stellar evolution, as well as disk evolution, planetary evolution, and the evolution of the overall stellar population (e.g., structure and kinematics). For example, if star formation is a rapid process, an association should have a small age spread \citep[e.g.,][]{Palla1997, Hartmann2001, Lee2005, Slesnick2008}, whereas if star formation occurs slowly, the observed age spread of a population should be large \citep[e.g.,][]{Feigelson1996}. The shape of the age distribution reflects the star formation history of the association: a peaked distribution indicates a possible initiating or concluding event for star formation \citep{Sancisi1974, DeGeus1992, Preibisch2002}, while a flatter distribution indicates a longer period of relatively constant star formation. Finally, the hierarchy of star formation is not clear: high mass stars may form either before or after low mass stars \citep{Elmegreen1977, Preibisch1999, Krumholz2019}. If there is an age gradient between different spectral types in an association, it could clarify the mass-dependent nature of the star formation process. 

It is difficult to accurately measure the ages of young stars, which are typically surrounded by large amounts of obscuring material and experiencing high levels of activity (e.g., accretion, star spots, and flares). The complex physics governing the collapse of a molecular core into a star is not fully understood, so evolutionary models have traditionally struggled to match observations \citep[e.g.,][]{Fang2013, Feiden2016, David2019}. Comparison to models is also difficult because a stellar ``zero age'' is difficult to define, so the earliest stages of a star's existence are poorly constrained: ultimately, all stellar age measurements are relative, rather than absolute. In associations, even those with a small age range, measurements are additionally complicated by the mass-dependent transition onto the Main Sequence (MS), which occurs progressively earlier for higher-mass stars \citep[e.g.,][]{Stahler1983, Hartmann1997}. After a star has transitioned to the MS, it becomes more difficult to measure the age, because over the MS lifetime a star moves minimally in \teff-luminosity space \citep[e.g.,][]{Chabrier1996}. All of these confounding factors make the important task of measuring the mean age and age spread of a young stellar association difficult.

The Scorpius-Centaurus OB association (Sco-Cen) is the nearest high-mass young stellar association, containing $\sim 150$ B-type stars \citep{deZeeuw1999}, with correspondingly large numbers ($\sim 10^{4}$) of low mass (G, K, and M) stars inferred by extrapolating the initial mass function (IMF; \citealt{Rizzuto2015}). Thus, Sco-Cen contains the majority of the young low-mass stellar population in the Solar neighborhood (distance $<$ 300 pc, age $<$ 20 Myr; \citealt{Preibisch2002}), which alongside its high-mass stellar population makes it an excellent laboratory for studying the outcomes of both low- and high-mass star formation. In general, the association is divided into three kinematically distinct groups: Upper Scorpius (US or Upper Sco), Lower Centaurus-Crux (LCC) and Upper Centaurus-Lupus (UCL) \citep{Blaauw1964, deZeeuw1999, Wright2018, Luhman2020}. Several studies \citep[e.g.,][]{Preibisch2002, Rizzuto2015, Luhman2020} have worked to determine membership of Sco-Cen, but these efforts have been complicated by the large spatial extent of the association on the sky ($\sim 80\degree \times 25\degree$; \citealt{Rizzuto2015}). 

Of the three main subgroups of Sco-Cen, LCC and UCL are both consistently estimated to be $\sim$17 Myr old \citep{Preibisch2008, Pecaut2012}. US appears to be younger, but its age remains much less certain than the age of LCC and UCL.  Some studies, especially earlier work \citep[e.g.,][]{DeGeus1989, deZeeuw1985,Preibisch1999, Preibisch2002} established the age as $\sim$5 Myr, while more recent results (e.g., \citealt{Pecaut2012, Pecaut2016}) have suggested that US has an age of $\sim$10 Myr. Surveys using different sub-populations of US have typically also found different ages, with low-mass (K and M) stars consistently appearing younger than the F and G type population \citep{Pecaut2012, Rizzuto2015, Pecaut2016}. This phenomenon has also been observed in other young populations, such as the Taurus star-forming region \citep{Gully-Santiago2017, Rizzuto2020} Several explanations have been proposed to explain this discrepancy, including star spots \citep{Rizzuto2015, Pecaut2016}, missing physics in models \citep{Feiden2016, David2019}, and a missing correction to account for F stars that have not yet reached the main sequence \citep{Fang2017}. 

There are additional factors in the uncertain age of US. Unresolved multiplicity has a twofold effect on observations of a system. First, the presence of a cooler secondary star can change the depth of temperature-sensitive spectral features like titanium oxide (TiO) bands in the red-optical regime, biasing the temperature measurement toward cooler temperatures than the true temperature of the primary. Second, unresolved binaries can be up to 0.75 mag more luminous than a single star of the primary's spectral type. This means that an unresolved binary will appear younger than a single star of its apparent spectral type would. These two effects combine to cause unresolved binaries to be measured as younger and less massive than the primary alone would be. These factors not only impact stellar measurements, but also change the interpretation of an observed population.


In addition to unresolved multiples, the large distance spread of US members (100-200pc, \citealt{Rizzuto2011, Rizzuto2015}) introduces distance uncertainty in measurements, particularly in surveys carried out before the \textit{Gaia} mission. For example, \citet{Preibisch1999} and \citet{Preibisch2002} assumed a distance of 145 pc for all their sources, because the 1 mas average \textit{Hipparcos} distance error was too large to distinguish any internal spatial structure. At a distance of 145 pc, 1 mas corresponds to a distance uncertainty of $\sim$ 25 pc, which is comparable to more recent measurements of the distance spread in Upper Sco ($\mu = 140$pc, $\sigma = 20$pc; \citealt{Rizzuto2015}). Distance uncertainty will not cause an age discrepancy as long as the mean age is known, but it can obscure observational signatures of multiplicity. For example, a 20 pc distance uncertainty can produce a 0.3 mag shift in an HR diagram at a distance of 140 pc, contributing to the blurring out of a cluster's sequence in an HR diagram and making binaries harder to detect.

Most corrections for multiplicity use a single value for the multiplicity fraction across all spectral types, which is typically either drawn from field star populations or from generalized studies of pre-main-sequence (PMS) multiplicity. However, the multiplicity fraction in PMS populations is generally about twice that of the field across all spectral types \citep{Ghez1997, Patience2003, Raghavan2010, Duchene2013}, and varies between star forming regions, with some regions, like Taurus and US, presenting high multiplicity fractions relative to the field (e.g., the companion fraction is $\sim$ 80\% for $0.25 M_{\odot} < M < 0.7 M_{\odot}$ in Taurus; \citealt{Kohler2000, Kraus2008, Duchene2013, Tokovinin2020}), while older populations like the Pleiades \citep{Bouvier1997} and denser regions like the Orion OB association \citep{Kohler2006, Reipurth2007} have multiplicity fractions that more closely resemble the field. For example, \citet{Tokovinin2020} measured the multiplicity in US for masses between 0.7 and 1.5 M$_{\odot}$ using speckle imaging, and found that the early M-dwarf multiplicity is $\sim 45\%$, which is approximately comparable to the G and K star multiplicity in Upper Sco. This differs from other star-forming regions and the field, where the abundance of multiples decreases to $\sim 30\%$ for M dwarfs \citep{Duchene2013, Ward-Duong2015, Winters2019}.

Other mass-dependent distributions are typically not considered at all in error corrections that rely on population synthesis, but the effect of these higher-order mass dependencies can be significant. For example, in addition to being region-dependent, the multiplicity fraction for a given star forming region is mass-dependent \citep{Kohler2006, Duchene2013}, along with other properties like the mass ratio and separation distributions. Differences in the shape of the mass ratio distribution can change the effect of multiplicity on measured population properties by increasing or decreasing the relative contribution of near-equal-mass vs. large-mass-ratio binaries. The mass-dependent properties of the binary population of a star-forming region, possibly coupled with the typical distance uncertainty to a star in US ($\sim 15-20$ pc; \citealt{Rizzuto2015}), may explain the apparent age gradient between FG and KM stars in US. Given that binary statistics such as multiplicity fraction, mass ratio distribution, and separation distribution are region- and mass-dependent, the statistics used to correct for the presence of binaries for a particular star forming region impact the interpretation of results.

Studies to measure the age of US have either not incorporated corrections for unresolved multiplicity \citep{Pecaut2012, Rizzuto2015} or have attempted to correct for multiplicity in a statistical manner \citep{Preibisch1999, Preibisch2002, Pecaut2016}. \citet{Preibisch1999} simulated a single-age, single-distance population of binary stars and estimated the average error multiplicity introduced into the luminosity of their simulated systems, but did not account for the shift in measured temperature caused by the presence of a binary. After combining the errors from their binary analysis with other observational errors, \citet{Preibisch1999} used their total observational error along with a stellar age model (at a set, preassigned age) to predict the range of HR diagram locations for a population of stars that matched the set stellar age they chose. The accuracy of \citet{Preibisch1999} method is dependent on their assumed binary population and star formation models and does not account for the mass dependence of binary properties.

Similarly, \citet{Pecaut2016} used multiplicity statistics for main sequence stars from \citet{Duchene2013} and \citet{Raghavan2010} to conduct a Monte Carlo analysis to account for the effects of multiplicity, observational uncertainties, spots, and intrinsic age spread. They retrieved HR diagram positions using evolutionary models, introduced multiplicity, and modified HR diagram positions based on the median observational uncertainties in their survey. They then constructed luminosity distributions from those positions, and compared their observed luminosity spreads to the simulated populations to identify the best fitting underlying age distribution after taking uncertainties into account. The assumptions made in this methodology include that a) observational effects are fully accounted for (and not suffering from their own biases), b) the underlying distributions used are representative of the true US population (e.g., that the multiplicity fraction is accurately reflected by the chosen value of 0.44, taken from \citet{Duchene2013} and \citet{Raghavan2010}, which does not account for mass dependence), c) measurement errors are not introduced, e.g., by inferring an effective temperature from a spectral fit, and d) binary properties, such as the mass ratio distribution, are not mass dependent. With some binaries inevitably remaining undetected in surveys, assumptions about the mass- and region-dependence of different binary population statistics like mass ratio distribution, separation distribution, and multiplicity fraction should alter measured age spreads, possibly in a mass-dependent manner.

To explore the effect of a population of undetected or uncorrected-for multiples on age measurements in Upper Scorpius, we have simulated a large stellar population with the same approximate statistics as US, including a more realistic representation of the low-mass binary population than has previously been considered. We created a population with a Gaussian age distribution, centered at 10 Myr with $\sigma =$ 2 Myr, which is consistent with many previous estimates of the age spread in US \citep{Preibisch2002, Preibisch2008, Slesnick2008, Pecaut2012}. We then performed a mock low-resolution red-optical spectroscopic survey of these populations, and found apparent effective temperatures for each system, treating each binary as an unresolved and undetected binary, and thus fitting each system with a single spectrum. By additionally propagating apparent magnitudes, and thus measuring an apparent luminosity for each system, we derived HR diagram positions for each system, and thus an age. Section \ref{sec:implementation} describes our population synthesis method in detail, while Section \ref{sec:fitting} describes our fitting algorithm for finding a best-fit effective temperature and extinction. Section \ref{sec:results} presents both the results from our initial tests and our full science results. We discuss our results in Section \ref{sec:disc}, and conclude in Section \ref{sec:conclusion}.

\section{Population Synthesis Method} \label{sec:implementation}
\subsection{Population Properties}

Our goal in simulating a population of stars was to replicate the observed properties of young stellar populations as closely as possible. We began our population synthesis by drawing a random mass for the primary star in each system using a modified \citet{Chabrier2003} initial mass function. The Chabrier IMF has been observed to be broadly consistent with observations in young star-forming regions \citep{Bastian2010}, and we modified it to reduce the number of low mass stars by a user-defined factor $f_{reduce}$ below a given cutoff mass $m_{cutoff}$. This choice was made to increase the relative number of high mass to low mass stars so that we could obtain statistically significant numbers of high-mass stars while keeping our computations efficient. For our typical sample size of 1200 systems, we produced $\sim$ 20-30 F stars before applying the reduction factor, which shifted to $\sim$ 200-300 F stars after applying the factor of $f_{cutoff} = 50$. Since studies of Upper Sco have typically focused on specific spectral types/mass ranges, rather than looking at the full population, we do not expect our results to be biased by this choice.

Since we are only examining stellar-mass companions, extending our primary star masses to the hydrogen-burning limit of 0.08 M$_{\odot}$ would artificially lower the number of companions for very low mass stars. To avoid suppressing the low-mass end of the companion mass distribution function too aggressively, we truncated our primary star mass function at 0.16 M$_{\odot}$. At the 0.16 M$_{\odot}$ lower limit for the primary star mass, 75\% of the secondary stars fall above the substellar mass limit, so the secondary star distribution at low masses is not significantly altered. A stellar mass of 0.16 M$_{\odot}$ also corresponds to \teff $\sim 3050$K, or a spectral type of M5 at an age of 10 Myr \citep{Pecaut2013}. The \citet{Pecaut2016} survey did not reach below 0.7 M$_{\odot}$, and \citet{Preibisch2002} had a lower mass limit of 0.1 M$_{\odot}$, so a mass cutoff of 0.16 M$_{\odot}$ does not reduce our ability to compare with either the \citet{Preibisch2002} or \citet{Pecaut2016} studies.

The Chabrier IMF has a log-normal form on its low-mass end, and follows a \citet{Salpeter1955} power law form on the higher-mass end, and thus has the functional form
\[
	\xi(\log(m)) \propto
	\begin{cases}
		\exp(\frac{(\log(m) - \log(m_{c})^{2}}{2  \sigma^{2}}) & \text{if } m \leq M_{cutoff}\\
		m^{-\alpha} & \text{if } m >  M_{cutoff}\\
	\end{cases}
\]

Thus, the relevant parameters for this IMF are the mean and standard deviation of the log-normal regime, the cutoff mass where the log-normal form changes to a power law, and the exponent of the power law. We used parameters from \citet{Chabrier2005}, which presents updated values from \citet{Chabrier2003}:
\begin{itemize}
\item Mean mass of lognormal distribution $m_{c} = 0.2 M_{\odot}$
\item Standard deviation of lognormal distribution $\sigma$ = 0.55 dex
\item Cutoff mass $M_{cutoff} = 1 M_{\odot}$
\item Salpeter power law slope $\alpha = -1.3$
\end{itemize}

To artificially suppress the number of low-mass systems, we separated out the relevant mass range of the IMF probability distribution function, divided it by the reduction factor $f_{cutoff}$, then appended it to the unaltered portion of the IMF, and proceeded with cumulative distribution function calculation and normalization as normal. This produced a low mass stellar distribution that still matched a Chabrier IMF, and a CDF with the appropriate normalization, but simply shifted the low mass portion of the CDF to become less likely to be drawn. The typical value of $f_{cutoff}$ was a factor of 50, and a typical mass cutoff point (below which the number of stars was reduced) was $m_{cutoff} = 0.8 M_{\odot}$. This cutoff mass is distinct from the cutoff mass of the Chabrier IMF, which transitions from a power law to a lognormal distribution at 1 M$_{\odot}$. The suppression mass was chosen so that it fell above the K-M spectral type transition so that any numerical effects caused by the truncation did not effect the population most susceptible to the effects of binaries. The suppression fraction was chosen to produce more comparable numbers of M and F stars when simulating a population. The populations above and below $m_{cutoff}$ both matched the functional form of the Chabrier IMF, simply with different normalizations, so the relative number of stars was changed but the final distribution of sources was not altered.

After assigning a mass to each primary star by drawing from the cumulative distribution function produced by the \citet{Chabrier2005} IMF, we assigned an age, a distance, and an extinction value to each system. All of these values were drawn randomly from an input distribution, assigned based on the physical parameters and star forming region we were interested in studying. For example, to replicate the distance distribution of systems in US, we used a Gaussian distance distribution centered at the mean distance of 140 pc, with a standard deviation of 20 pc \citep{Rizzuto2015}. The extinction in US is low, generally ranging from $0 \leq A_{V} \leq 2$ \citep{Preibisch2002}, and each system was assigned a value drawn uniformly within those limits. 

Many different functional forms of the shape of the star-formation rate over time have been suggested: a ``burst'' scenario, where there is a minimal age spread \citep[e.g.,][]{Slesnick2008, Preibisch2008}, a Gaussian distribution \citep[e.g.,]{DaRio2010, Pecaut2012}, and linearly or exponentially increasing or decreasing with time \citep[e.g.,][]{Dahm2008, Fang2013}. \citet{Pecaut2012} found a Gaussian distribution of ages with spreads of $\sigma = $3-5 Myr (observed), which they convert to an intrinsic age spread of $\sigma =$ 2-6 Myr, which are upper limits because they do not correct for multiplicity. This narrow age spread is consistent with other US age spread measurements \citep[e.g.,][]{Preibisch2002, Preibisch2008, Slesnick2008}. \citet{Pecaut2016} observed an age spread of $\sigma =$ 7 Myr using low-mass stars, which they then discarded, instead using the Main-Sequence G star turn-off for age measurements, which resulted in an observed age spread of $\sigma =$ 7 Myr after performing a Monte Carlo correction for observational uncertainty. However, that correction still has many associated uncertainties (as discussed in Section \ref{sec:intro}). Thus, we prefer the earlier results, which are presented as upper limits without an attempt to correct for multiplicity, and adopt a linear Gaussian age distribution with $\mu = 10$ Myr and $\sigma = 2$ Myr, which is within the ranges presented by much of the previous literature. Future work may consider the effects of different star formation histories on our results.

\subsection{Binary Statistics}\label{sec:binstat}

\begin{deluxetable*}{cCCCcCcCc}
\tablecaption{Summary of data used to infer mass-dependent models \label{tab:multiplicity stats}}
\tablecolumns{9}
\tablewidth{0pt}
\tablehead{
\colhead{Spectral type} &
\colhead{Approx. mass \tablenotemark{a}} &
\colhead{$\mu_{\log_{10}(a)}$} & \colhead{$\sigma_{\log_{10}(a)}$} & \colhead{Ref.}
& \colhead{$\Gamma$\tablenotemark{b}} & \colhead{Ref.} & \colhead{Multiplicity} & \colhead{Ref.}\\
\colhead{} & \colhead{(M$_{\odot}$)} & \colhead{($\log(AU)$)} & \colhead{($\log(AU)$)} & \colhead{} & \colhead{} & \colhead{} & \colhead{} & \colhead{}
}
\startdata
A0 & 2 & 2.59\pm0.13 & 0.79\pm0.12 & (1) & -2.3$^{+1.0}_{-0.9}$ & (1) & 70\% & (1)\\
G0 & 1 & 1.65\pm 0.5\tablenotemark{c} & 1\pm0.2\tablenotemark{c} & (2) & 0\pm0.5\tablenotemark{c} & (2) & 45\% & (2)\\
M0 & 0.6 & 1.69$^{+1.5}_{-1.0}$& 1\pm0.1 & (3,4) & 0.18$^{+0.33}_{-0.30}$ & (3) & 45\% & (5)\\
M4 & 0.2 & 1$^{+0.7}_{-0.33}$ & 0.8\pm0.2 & (3,4) & 1.02$^{+0.59}_{-0.52}$ & (3) & 35\% & (3)\\
M9 & 0.1 & 0.65$^{+0.16}_{-0.11}$ & 0.6\pm0.1 & (3) & 4$^{+1.9}_{-1.6}$ & (3) & 25\% & (3)\\
\enddata
\tablenotetext{a}{Spectral type to mass conversion (at field age) from \citet{Pecaut2013} and the MIST evolutionary tracks}
\tablenotetext{b}{Power-law index for the mass ratio distribution}
\tablenotetext{c}{Errors are typical errors for each measurement, but errors for these measurements in the papers are not given.}
\tablecomments{(1) \citet{DeRosa2014}; (2) \citet{Raghavan2010}; (3) \citet{Kraus2012}; (4) \citet{Winters2019}; (5) \citet{Tokovinin2020}}
\end{deluxetable*}

The properties of binary star populations - the mean and standard deviation of the separation distribution, the mass ratio distribution, and the multiplicity fraction - are all mass dependent. To parameterize these distributions, we used statistics for populations of stars across the mass range of our study (0.16 M$_{\odot} \lesssim$ M $\lesssim$ 2 M$_{\odot}$) to derive functional forms of distributions to draw from. We used statistics corresponding to field star spectral types of A0, G0, M0, M4, and M9, roughly equivalent to stellar masses of 2, 1, 0.6, 0.2, and 0.1 M$_{\odot}$. For A stars, we referred to \citet{DeRosa2014}, for G stars we used the statistics of \citet{Raghavan2010}, and for M stars we used a combination of statistics from \citet{Kraus2012}, \citet{Winters2019}, and \citet{Tokovinin2020}. 

The multiplicity fraction of stars is mass dependent \citep{Duquennoy1991, Raghavan2010, Duchene2013}, which is relevant given the wide mass range of this study (0.1 M$_{\odot} \lesssim$ M $\lesssim$ 2 M$_{\odot}$). We randomly assigned primary stars to be either single or binary using a mass-dependent multiplicity fraction. Each multiplicity fraction, along with its corresponding mass range and relevant references, is listed in Table \ref{tab:multiplicity stats}. The sources for the multiplicity fraction in each mass range used different methods for binary discovery, and so had different sensitivity limits in both mass ratio and separation. However, in regimes where there was overlap between different sources, we compared results and found that different methods typically found similar values (e.g., \citet{Kraus2012} and \citet{Winters2019} for M stars).

Because any survey of multiplicity is inevitably incomplete at some level, the measured multiplicity fractions are technically lower limits on multiplicity. However, these fractions are probably close to the true value, because surveys are typically sensitive to well inside the peak of the semimajor axis distribution, and across that separation range they are sensitive to masses down to near the substellar boundary. Therefore our binary fractions may slightly underestimate the true binary fraction in each mass range. Although our overall binary fractions may be underestimates, it is possible that our binary population could be overestimates in certain parts of the mass ratio-separation parameter space because of the varying detection limits and methodologies between different surveys. Another potential cause of an overestimate would be if multiplicity surveys used corrections for targets outside their sensitivity range (e.g., in the low-mass-ratio regime for high-mass targets) that differed from the assumptions we made in constructing our binary population. However, this second-order effect would be difficult to account for, and if it is present it should be small. Therefore, we may be slightly over- or underestimating the number of binaries, but we do not expect the effect to be large. We assigned multiplicity using mass bins centered in between the points defined in Table \ref{tab:multiplicity stats}, rather than fitting a function.  

\begin{figure}
\gridline{\fig{sep.pdf}{0.95\linewidth}{}}
\gridline{
	\fig{sep_stdev.pdf}{0.95\linewidth}{}}
\gridline{
	\fig{gamma.pdf}{0.95\linewidth}{}}
\caption{Data from Table \ref{tab:multiplicity stats} plotted with models. The red error bars are values estimated from typical errors in the relevant measurements, because the literature did not list an error value. \textbf{Top: }The mean of the lognormal separation distribution is approximately linear in log(mass). \textbf{Middle: }The standard deviation of the lognormal separation distribution is approximately quadratic in log(mass). \textbf{Bottom:} The mass ratio power law exponent was determined using a piecewise linear interpolation}
\label{fig:statplot}
\end{figure}

The mass ratio distribution is typically fit as a power law, with an index that is mass-dependent \citep{Burgasser2006, Raghavan2010, Kraus2012, DeRosa2014}. We modeled the mass ratio power law index using a piecewise linear interpolation based on the data presented in Table \ref{tab:multiplicity stats}, with the space between each set of points being modeled by a line. We chose to interpolate between the points to create a more smoothly varying distribution for $\Gamma$, which we assume more closely mimics the underlying mass dependence of $\Gamma$. A plot of the data points and the resulting piecewise linear interpolation is shown in Figure \ref{fig:statplot}. We truncated the mass ratio distribution at a value of 0.1, and the minimum mass of a secondary star to be 0.1 M$_{\odot}$, meaning that at $M \leq 1 M_{\odot}$ the minimum mass was 0.1 M$_{\odot}$, and between 1 and 2 M$_{\odot}$ the minimum mass increased to 0.2 M$_{\odot}$. This was to avoid the effect of the negatively-sloped mass ratio distribution for high-mass stars blowing up at very small mass ratios, and to restrict the companion masses to above the substellar boundary.

The separation distribution is well modeled by a lognormal distribution \citep[e.g.,][]{Raghavan2010, Duchene2013, DeRosa2014, Winters2019}, and the mean and standard deviation of that distribution are mass dependent. We modeled the mean and standard deviation of the separation distribution as 
\begin{equation}
\begin{split}
\mu_{\log(a)} &= \alpha \log(m) + \beta\\
\sigma_{\log(a)} &= \xi(\log(m) - \theta)^{\kappa} + \lambda
\end{split}
\end{equation}
where for $\mu_{\log(a)}$, $\alpha = 1.48 \pm 0.08$ and $\beta = 2.12 \pm 0.06$, and for $\sigma_{\log(a)}$ $\xi = -0.93 \pm 0.06$, $\theta = -0.18 \pm 0.02$, $\kappa = 2$, and $\lambda = 1.02 \pm 0.01$, using a least-squares fit to the data of Table \ref{tab:multiplicity stats}. Figure \ref{fig:statplot} shows plots of the data and models for both the mean and standard deviation of the separation distribution as function of primary star mass. \citet{Raghavan2010} presented a semimajor axis distribution, but the other sources for separation distributions did not correct for projection effects, instead calculating a projected separation distribution. We treated all separation distributions as projected separations. The distribution of projected separations is slightly broader than that of semimajor axes (a result of convolving the semimajor axis distribution with the distribution of orbital positions for each system), so the width of the distribution for G stars in our model is slightly smaller than that of the resultant projected separation distribution.

\citet{Moe2017} found that the shape of the separation distribution is dependent on mass ratio for A stars, and \citet{DeRosa2014} also found that close A star binaries likely have a more uniform mass ratio distribution than those with wide separations. However, we choose to still model the A star mass ratio distribution as a single lognormal distribution for two reasons: First, the \citet{Moe2017} sample of A stars spans a mass range of $2 < M < 5 M_{\odot}$, but 2 $M_{\odot}$ is the maximum mass in our synthetic population. At spectral types below A stars, the mass ratio distribution does not appear to be separation dependent \citep{Raghavan2010, Moe2017}, so we would expect the separation dependence to only be relevant to our highest-mass systems. Second, the separation distribution for A stars peaks at log(a) = 2.59, which corresponds to log(P) $\sim$ 4. At log(P) = 4, the mass ratio power law index is between -1 and -2. This means that the majority of the A star population falls in the steep-mass-ratio/large separation regime. Therefore, we model the separation distribution as a single lognormal that is not dependent on mass ratio, because we would expect the effect of that dependence on our results to be small. 

\begin{figure}[t]
\plotone{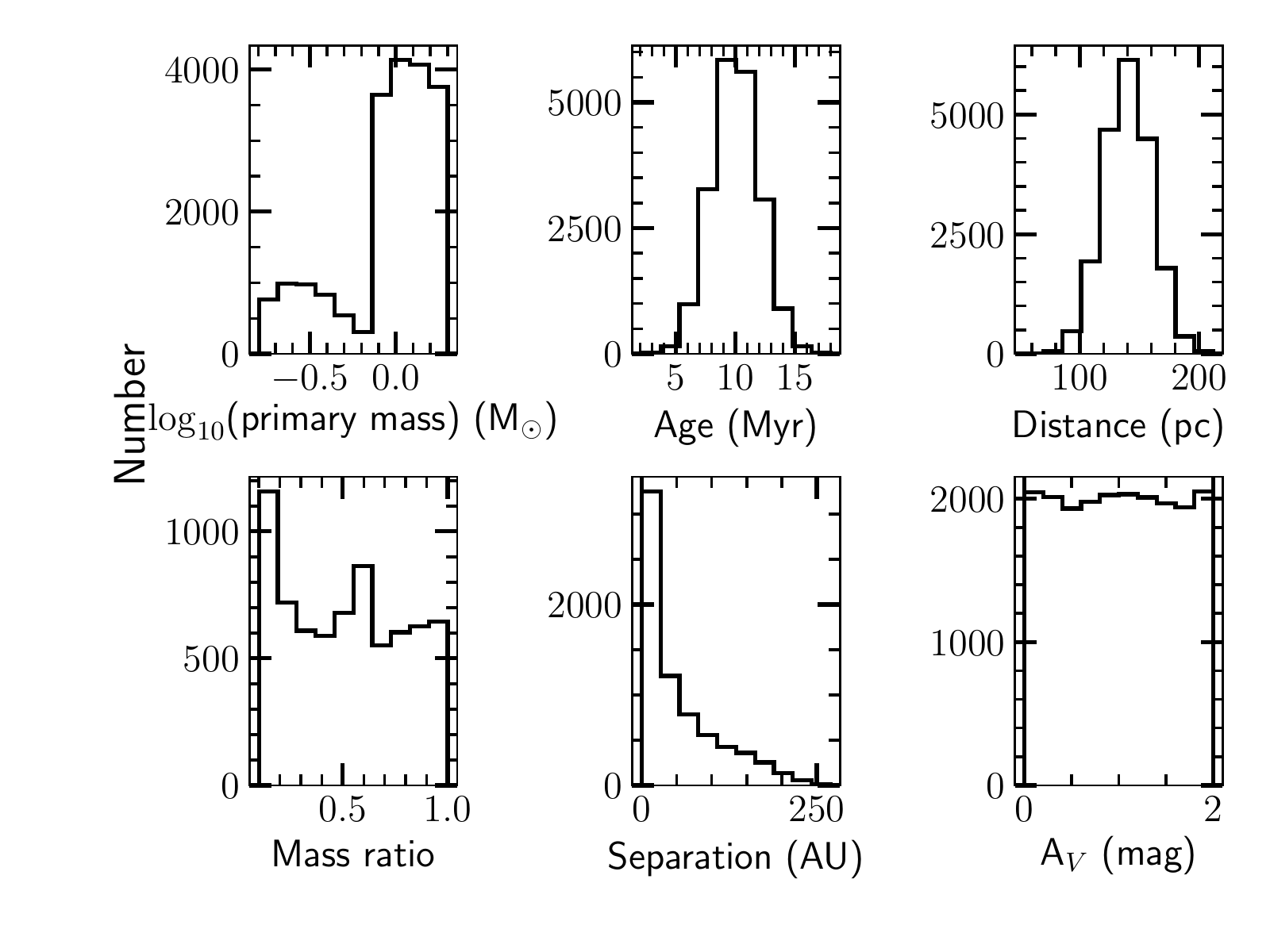}
\caption{Distributions of input parameters for primary star mass, age, distance, mass ratio, separation, and extinction. The apparent bimodality of the primary star mass distribution is because the low mass stars have been suppressed, causing the low-mass portion of the primary star mass function to sharply decrease relative to the high-mass end. This method was implemented so that our simulation produced roughly similar numbers of M and F stars. The suppression of low-mass stars also causes high-mass stars to dominate the mass ratio distribution, causing a peak toward zero due to their very negative power law exponent. The peak at q = 0.5 is a result of not allowing the secondary star mass to go below 0.1 M$_{\odot}$, so low-mass stars are skewed toward higher mass ratios.}
\label{fig:pars}
\end{figure}

We assigned each binary system a mass ratio and separation using these mass-dependent distributions. We then filtered out systems where the secondary star would likely fall outside the spectrograph slit in a survey by ignoring the assigned secondary star for systems with an angular separation $> 1.5\arcsec$, which we calculated using the assigned system distance and separation. We chose a value of 1.5$\arcsec$ because it falls between the 1$\arcsec$ slit size used by \citet{Pecaut2012}, \citet{Rizzuto2015}, and  \citet{Pecaut2016}, and the 2$\arcsec$ diameter fiber used by \citet{Preibisch1999} and \citet{Preibisch2002}. It is also comparable to good seeing in spectroscopic conditions at most observing sites. 

For systems that would have been resolved, we neglected the secondary star and only considered the primary. Thus, our final set of systems only contained binaries that would be unresolved in a typical survey with angular resolution of 1.5$\arcsec$. At the average US distance of 140 pc, this corresponds to a maximum physical separation for unresolved binaries of $\sim$210 AU. Figure \ref{fig:pars} shows the input distributions for primary star mass, age, distance, mass ratio, separation, and extinction for our final code run mimicking Upper Sco, which consisted of 1200 systems with a mean age of 10 Myr (RMS = 2 Myr). The number of low-mass stars was suppressed, causing a sharp change in the number of stars around 0.8 M$_{\odot}$. 

\subsection{Physical System Parameters}
We used the the MESA Isochrones and Stellar Tracks (MIST) stellar evolutionary models \citep{Paxton2011, Paxton2013, Paxton2015, Dotter2016, Choi2016} to derive observable parameters from the assigned age and mass for each star in each system. We chose to use MIST models instead of the Baraffe or Dartmouth models \citep{Baraffe2015, Feiden2015} because the MIST models span broader mass and age ranges, and are more finely spaced than the PARSEC models \citep{Bressan2012}. Although each of these models has different methodology and physics, we expect our results to be model-invariant, at least to the first order, because our interest is in differential, rather than absolute, measurements.

For a given age and mass, the evolutionary models predict the luminosity, log(g), temperature, and UBVRIJHK photometric absolute magnitudes, allowing us to convert from physical parameters to observables. For each output term of interest, we bilinearly interpolated the values given by the evolutionary models as a function of age and mass. For each star in each system, given the randomly chosen mass ratio, primary mass, age, and distance, we used the MIST models to infer effective temperature (\teff), bolometric luminosity (L), and VRIJHK magnitudes.

We assumed all systems had solar metallicity, because the metallicity of most nearby star-forming regions is approximately solar \citep{James2006, Biazzo2011, D'Orazi2011}, although future work may explore the effects of metallicity on our results. Additionally, star-forming regions are expected to have a roughly uniform abundance \citep[e.g.,][]{Hawkins2020}, so changing the single-valued metallicity of the full simulated population should not affect our results significantly.

\begin{table*}
\caption{Input and derived system parameters for our model.}
\begin{tabular}{cccc}
\hline
 \multicolumn{4}{c}{System Parameters}\\
\hline
Parameter & Fit? & Range & Description\\
 \hline
Mass & Y & 0.16 - 2 M$_{\odot}$  & Primary star mass \\
Age & Y & Gaussian with $\mu = 10$Myr, $\sigma = 2$Myr & Age range\\
Distance & N & Gaussian with $\mu = $140 pc, $\sigma = 20$ pc & Distance to each system\\
Multiplicity & N & 0 or 1 & Drawn using mass dependent multiplicity fraction\\
Mass ratio* & N & Power law & Drawn using mass dependent power law\\
Projected separation*  & N & Lognormal & Drawn using mass dependent separation distribution\\
Temperature$^{+}$ & Y & \nodata & Drawn from isochrone \\
$\log(g)^{+}$  & N &  \nodata & Drawn from isochrone \\
Extinction & Y & A$_{V}$ = 0-2 mag (uniform) & Randomly selected for each system\\
Luminosity $^{+}$ & Y & \nodata & Bolometric luminosity\\
VRIJHK magnitude$^{+}$  & N & \nodata & Johnson-Cousins filters\\
 \hline
\end{tabular}
\tablecomments{An asterisk denotes a value that is only assigned if a system is a binary, and a $^{+}$ denotes a parameter that is derived from models based on other, assigned values. Parameters with a Y in the ``Fit?'' column had output values determined by the fitting process, while an N indicates that the value was held constant or was calculated independently from the fit.}
\label{tab:params}
\end{table*}

To create observed photometry for each system, we converted each system's assigned A$_{V}$ to A$_{RIJHK}$ extinctions using the conversions given in \citet{Cardelli1989}. To obtain absolute magnitudes, we converted the model photometric magnitudes into flux space before interpolating them, and assumed that the parameters were linear between the grid points of the evolutionary models, which are spaced at $\sim 0.04 M_{\odot}$ in mass and 0.05 Myr in log(age). We used the absolute magnitudes along with our assigned system distance and system extinction to calculate apparent magnitudes for each system. For binaries, we converted the two magnitudes to fluxes, summed them, then converted back to a system magnitude before calculating apparent magnitudes. Table \ref{tab:params} shows a summary of the various input parameters during system creation.

\subsection{Overview of Spectral Analysis Methods Applied to Upper Sco Members}
Our eventual goal was to compare our simulation results to the various studies of stellar members of Upper Sco from the last two decades: \citet{Preibisch2002}, \citet{Pecaut2012}, \citet{Rizzuto2015}, and \citet{Pecaut2016}. Therefore, we selected wavelength ranges, resolutions, spectral typing techniques, and extinction measurement methods that resembled the observational and analytical methods used by those various studies. However, the techniques used for spectral typing and extinction measurements are not always consistent between studies. Extinction is typically measured using either a simultaneous fit to temperature and A$_{\lambda}$, or by establishing spectral type or temperature using a continuum-flattened spectrum, then measuring extinction from the color excess by comparing the observed colors to some set of intrinsic colors. Flat fielding spectra is often complicated and subject to the design characteristics of the instrument used in the observations, and so the simultaneous fit technique is often not possible, as it requires the true continuum of a flat-fielded spectrum. Appendix \ref{sec:appendixA} discusses these two techniques further.

Different methods may be used to spectral type stars. Typically either synthetic spectra from model atmospheres or empirical spectral type templates are used, with the latter either coming from a set of young stars or field stars. Model spectra are not perfect matches to observed spectra (due to, e.g., incomplete line lists), but they do provide a direct connection from spectral features to a temperature. Empirical templates of young stars will match other young stars well, but there is uncertainty in the mapping between spectral type and temperature \citep[e.g.,][]{Luhman2003, Pecaut2013, Herczeg2014}. The connection between spectral type and temperature for field star templates may be better calibrated than those of young stars, but the mismatch between young and Main Sequence stars introduces additional uncertainty beyond the intrinsic error of converting between spectral type and temperature scales. 

\citet{Preibisch2002} observed their Upper Sco sample using the 2dF multi-object spectrograph at the Anglo-Australian Telescope. The observations spanned a wavelength range of $6150 < \lambda < 7250$\AA, with a resolution of 2\AA. With a central wavelength of 6700\AA, this corresponds to R$\sim$3350 (R $\equiv \lambda_{cen}/\Delta \lambda$). They measured the spectral type of their stars by comparing to empirical templates of other Upper Sco stars (compiled from \citealt{Walter1994, Kohler2000, Martin1998, Ardila2000}) after dividing out the continuum. They measured extinction by calculating the R-I color excess using an intrinsic color-vs-temperature relation from \citet{Hartigan1994}, then converted E(R-I) to A$_{V}$ using a relation from \citet{Herbig1998}, and dereddened assuming that $A_{I}$ = 0.482 A$_{V}$ \citep{Rieke1985}. 

\citet{Preibisch2002} found a small error of only 0.3 spectral type subclasses, but the use of empirical templates requires a conversion from spectral type to temperature using evolutionary models, which introduces error into the temperature measurement. Measuring extinction with just one color gives no information on the possible error on the extinction measurement, and extinctions derived from colors are only as accurate as the extinction curve and intrinsic color measurements used.

\citet{Pecaut2012} and \citet{Pecaut2016} both took their observations using a spectrograph at the SMARTS telescope at Cerro Tololo Inter-American Observatory. Their spectra had blue and red arms spanning $3700 < \lambda < 5200$\AA\ and $5600 < \lambda < 6900$\AA, respectively, and resolutions of R$\sim$1100 (blue; at H$\beta$) and R$\sim$2100 (red; at H$\alpha$). \citet{Pecaut2012} used \textit{V} magnitudes and \textit{B-V} and \textit{V-I$_{C}$} colors from the \textit{Hipparcos} catalog \citep{Perryman1997} and \textit{JHK$_{S}$} photometry from 2MASS \citep{Skrutskie2006}.  \citet{Pecaut2016} used these catalogs, and also obtained \textit{BV} photometry from Tycho-2 \citep{Hog2000}, APASS \citep{Henden2012}, and SACY \citep{Torres2006} surveys. 

Both \citet{Pecaut2012} and \citet{Pecaut2016} classified their stellar spectra using an empirical grid of spectral standard stars after removing the continuum with a spline fit. \citet{Pecaut2012} calculated extinction using an intrinsic temperature-color relation compiled from various sources in the literature \citep{Perryman1997, Mermilliod2006, Stauffer2010}. They calculated five color excesses: $E(B-V), E(V-I_{C}), E(V-J), E(V-H)$, and $E(V-K_{S})$, and used extinction ratios from \citet{Fiorucci2003} to convert to A$_{V}$. They used the median A$_{V}$ as the ``true'' value, and took the standard deviation of the five values as an estimate of the error on the measurement. \citet{Pecaut2016} used the same method, but used intrinsic color-temperature relations from \citet{Pecaut2013} and excluded $E(V-I_{C})$ from their calculation. The same spectral typing caveats from the \citet{Preibisch2002} paper are true here, but the extinction measurement is more rigorous because there are multiple colors used. However, extinctions derived from colors rely on other measurements that have some error associated with them.

\citet{Rizzuto2015} observed their sample of low-mass Upper Sco stars using the Wide-Field Spectrograph on the Australian National University 2.2m telescope. Their observations spanned a blue and a red arm of the spectrograph, wth R$\sim$3000 in the blue end ($3600 < \lambda < 4800$\AA) and R$\sim$7000 at the red end (4800 or 5300\AA$< \lambda < $7000\AA). To spectral type and fit for extinction simultaneously, \citet{Rizzuto2015} did not continuum normalize their spectra, and instead fit for \teff\ and A$_{V}$ simultaneously using a grid of empirical templates \citep{Pickles1998, Bochanski2007} and the \citet{Savage1979} extinction law. This method still uses empirical templates, but fitting for extinction and temperature simultaneously reduces the systematic errors for both values relative to if they had been measured independently.

All of these surveys used similar resolutions and wavelength ranges. We chose to use a wavelength range of $\lambda = 5600-6900$ \AA\ and a resolution of $R \sim 2100$, to match the \citet{Pecaut2012, Pecaut2016} observations while still falling in or close to the wavelength range of \citet{Preibisch2002} and \citet{Rizzuto2015}. We used the lowest resolution for computational efficiency, but found that using different resolutions and wavelength ranges did not significantly impact the quality of our fits. We adopted the simultaneous \teff-A$_{V}$ fitting method of \citet{Rizzuto2015}, but explored the effects of using the simultaneous method versus the asynchronous photometric method in Appendix \ref{sec:appendixA}. We found that the typical difference between the two methods was $< 20$ K in temperature and $< 0.1$ mag in A$_{V}$.

\subsection{Spectrum Creation}\label{sec:make spec}
\begin{figure*}
\plotone{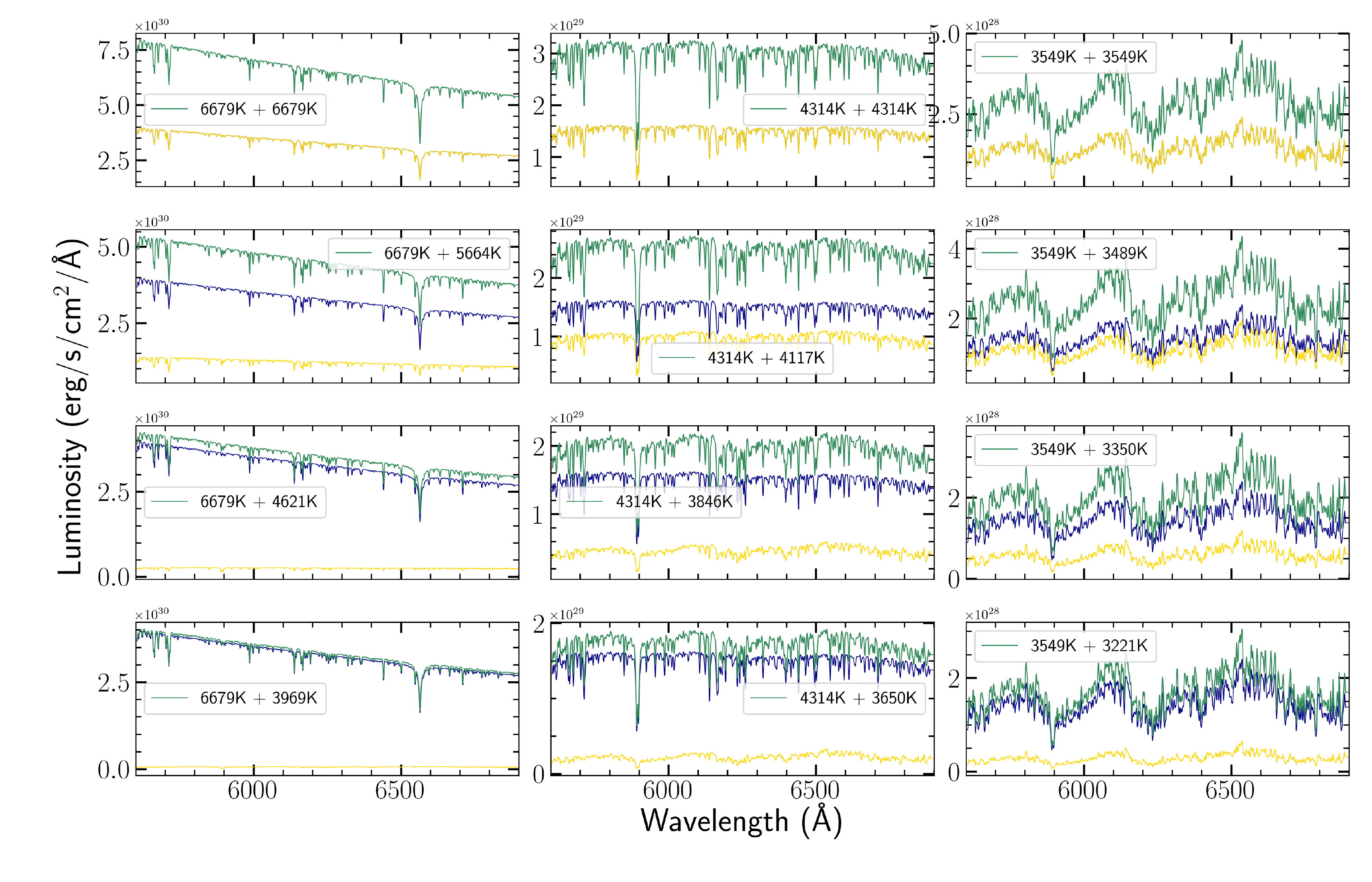}
\caption{An F2 (left), K4 (center), and M2 (right) primary star with secondary stars with mass ratios of 1, 0.9, 0.7, and 0.5 (from top to bottom). The spectral type to temperature conversion was made using \citet{Pecaut2013}, and the temperature to mass conversion was made using the \citet{Dotter2016} (MIST) models. The composite spectrum is shown in green, the primary in dark blue, and the secondary in gold. For a mass ratio of 1, the composite spectrum is identical to the stellar spectrum, but the luminosity is doubled. As the mass ratio decreases, the contribution of the secondary begins to have an effect on the spectral morphology, but at small mass ratios the effect of the secondary is not noticeable due to the large luminosity difference between the two components. If one of the components is not visible in the figure it is because it is overlaid with another spectrum (e.g., in the top panel, the primary and secondary spectra are overlaid, so only the secondary is visible). The effect of multiplicity is more pronounced for later spectral types. Note the differing y-scales between panels.}
\label{fig:example specs}
\end{figure*}

After deriving system observables from isochrones, we produced a spectrum of each star in each system. We began with the \citet{Husser2013} high-resolution spectra computed with the PHOENIX stellar atmosphere code, which we chose for their large range of \teff\ and log(g) values. The model spectra were gridded in \teff\ and log(g), so we selected the four models closest to each star's \teff\ and log(g) and bilinearly interpolated between them to obtain a spectrum for each primary and secondary star. 

To best mimic the low-resolution optical spectra of previous studies of US, we simulated spectra with R$\sim 2100$ spanning from 5600 - 6900 \AA, matching the observations of \citet{Pecaut2016}. Other surveys \citep[e.g.,][]{Preibisch2002, Rizzuto2015} used similar resolutions and wavelength ranges. We chose to match \citet{Pecaut2016} because we wished to investigate the effect multiplicity would have on an older population, like that suggested by \citet{Pecaut2016} and \citet{Pecaut2012}. We created the requisite ``instrumental'' resolution by convolving the spectrum with a Gaussian distribution, then imposed pixelation of 3 pixels per resolution element by binning the spectrum, to mimic the pixelation of a detector. To determine the parameters of the convolution kernel, we set the full width at half maximum (FWHM) of the Gaussian to be the mean wavelength of the spectrum, divided by the resolution. 

We converted each stellar spectrum to physical units of flux at the stellar surface (erg/s/cm$^2$/\AA) by noting that $L = 4\pi r^{2} \sigma T^{4} \implies r = \sqrt{\frac{L}{4\pi \sigma T^{4}}}$, where L and T are the luminosity and temperature assigned to a given mass and age by the evolutionary models, and $\sigma$ is the Stefan-Boltzmann constant. Then, the flux at the stellar surface is simply the surface area of the star times the unit stellar flux in the spectrum. To apply extinction, we used the extinction curve defined in \citet{Cardelli1989}, which assumes an R$_{V}$ of 3.1. 

\begin{table*}
\begin{center}
\caption{Luminosity and flux ratios for example systems and bands}
\begin{tabular}{ccccccccccccc}
\hline
Mass ratio & \multicolumn{3}{c}{L$_{bol}$ ratio} & \multicolumn{3}{c}{V ratio} & \multicolumn{3}{c}{I$_{c}$ ratio} & \multicolumn{3}{c}{K ratio}\\
\hline
& F2 &K4 & M2 & F2 &K4 & M2 & F2 &K4 & M2 & F2 &K4 & M2\\
 \hline
0.9 & 0.36 & 0.74 & 0.86 & 0.32 & 0.41 & 0.56 & 0.59 & 0.69 & 0.81 & 0.72 & 0.84 & 0.89 \\
0.7 & 0.09 & 0.46 & 0.58 & 0.02 & 0.03 & 0.08 & 0.29 & 0.44 & 0.63 & 0.39 & 0.58 & 0.67\\
0.5 & 0.03 & 0.28 & 0.36 & 3.5$\times 10^{-3}$ & 0.01 & 0.04 & 0.14 & 0.29 & 0.49 & 0.20 & 0.39 & 0.50\\
 \hline
\end{tabular}
\label{tab:lum ratio}
\end{center}
\end{table*}

Our goal was to fit both single and binary stars as if they were single stars, thus simulating a population containing undetected binaries. If a system was a binary we summed the two component spectra to produce an unresolved system spectrum. Figure \ref{fig:example specs} shows an example of F2, K4, and M2 primaries with secondary stars determined by mass ratios of 1, 0.9, 0.7, and 0.5. The secondary star with equal mass ratio doubles the observed luminosity without affecting the spectrum, then the contribution decreases until it becomes insignificant at small mass ratios. The specific luminosity ratio for each spectral type and mass ratio combination is listed in Table \ref{tab:lum ratio}. We tested adding synthetic noise to the spectrum at a level equivalent to a signal-to-noise ratio of 50, but found that it dramatically slowed down the speed of our fit without significantly changing our results, so we did not add noise to the generated spectra. Our final output was the physical and observable properties of each system, including orbital elements and system parameters for each binary, and a single spectrum for each system.

\section{Model Fitting and Parameter Retrieval Method}\label{sec:fitting}

To analyze our synthetic population, we matched each output model system spectrum (hereafter referred to as our data) with a best-fit model spectrum. We wished to model the fitting procedures typically used by observers, and so used the \citet{Husser2013} model spectra calculated with PHOENIX stellar atmosphere models to perform a least-squares fit to our simulated data, interpolating the spectra as described in Section \ref{sec:make spec} to produce finer \teff\ resolution than the 100K resolution of the models. 

We used a fitting algorithm that mimicked a Gibbs sampler (a common sampler used in Markov Chain Monte Carlo techniques), but that only moved downward in $\chi^{2}$ space, in contrast to a traditional Gibbs sampler, which randomly may decide to step upward in $\chi^{2}$ space. The algorithm starts in an arbitrary location, then randomly varies the fit parameters, accepting tries that reduce the $\chi^{2}$ value of the fit. We chose this technique over a simpler method (such as an amoeba algorithm) to reduce the effects of a potentially quite complicated $\chi^{2}$ surface. Given that the calculation is based on the interpolation of grids of model spectra, we also wished to avoid the complexity of differentiation, which is required for more sophisticated gradient-descent techniques.

We fitted for three parameters: \teff, extinction, and normalization factor. The normalization factor was a multiplicative factor that scaled the test spectrum so that it roughly matched the luminosity of the data spectrum, so that we did not make assumptions about, e.g., stellar radius, that would be necessary without an arbitrary normalization factor. We tested adding log(g) as an additional fit parameter, but found that it significantly slowed down the fit and was not well constrained. The typical scatter in our log(g) recovery was 0.2 dex, which is comparable to the maximum difference in log(g) between the two ends of our mass range ($\sim$0.08-2 M$_{\odot}$) at an age of 10 Myr. This meant that any difference in log(g) measurement caused by the presence of a binary would not have been resolved by our log(g) fits. Therefore, we held log(g) constant at the correct value for the primary star during the fitting process.

We created an initial guess for each parameter (\teff, normalization, and extinction), which was the initial value for each system plus some normally distributed perturbation. We chose each initial value independently to avoid imposing initial assumptions. For example, A$_{V}$ and \teff\ were varied independently to avoid the natural covariance of those two parameters, meaning that the fit could move more rapidly to better values. For the first fitting stage, the \teff\ initial guess was the primary star's temperature modified by a value drawn from a normal distribution with a standard deviation of 100 K. In A$_{V}$, the initial guess was the assigned A$_{V}$ plus a value drawn from a normal distribution with a standard deviation of 0.1 magnitudes. In normalization, the initial guess was the maximum value of the test spectrum divided by the maximum value of the data spectrum, perturbed by a normal distribution with a standard deviation that was 2\% of the normalization value. The size of perturbation and initial guesses were chosen to try to mimic the approximate accuracy an observer might achieve with a by-eye inspection of a spectrum. 

We fit our data using two stages, both of which used a low-resolution (R $\sim$ 2100) spectrum spanning 5600 - 6900 \AA. We ran the first stage briefly to determine an initial set of parameters, assuming that if the initial guess was very far offset from the best fit values it would approach a better value quickly. This also served the purpose of debiasing our initial guess, serving much like a burn-in time would for an MCMC simulation. Using the best fit parameters from this initial fitting stage, we continued fitting for a longer time to ensure that we reached convergence.

In each stage, after making our initial guesses, for each test step we varied the normalization, \teff, and extinction simultaneously. For the first $N_{steps}/2$ steps, we varied the parameters using coarse parameters drawn from a normal distribution with widths 100 K, 10\%, and 0.05 mag for \teff, normalization, and extinction, respectively. For the second $N_{steps}/2$ steps, we used perturbations drawn from normal distributions with widths of 5K, 0.5\%, and 0.01 mag in \teff, normalization, and extinction. 

We chose our step sizes based on two considerations. For our initial fitting step, we wished to achieve an accuracy in \teff\ of approximately 1 spectral subclass. For an M star, the difference between subclasses is $\sim$ 100 K \citep[e.g.,][]{Herczeg2014}, meaning that our initial step size in \teff\ had a variance of 100K. Then, we noted that the normalization value is a proxy for the luminosity, meaning that \teff\ and normalization are related as $L \propto T^{4}$, so changes to the temperature change the luminosity. For a 4000K star, a 100K change is $\sim$ 2.5\% of the total value, and changes the luminosity by $1.025^{4} = 1.10$, or 10\%. We wished to explore the parameter space to the same extent in all three directions, meaning that in normalization and A$_{V}$ we needed to produce variations of $\sim$ 10\%. Thus, our normalization variance for the large steps was also 10\%. For A$_{V}$, the effect of 0.1 mag of A$_{V}$ produces $\sim$ a 10\% change to the flux in the optical regime. We found that our fit was very sensitive to A$_{V}$, and so reduced the variance from 0.1 to 0.05 mag to achieve the highest accuracy possible.

After coarsely exploring the parameter space, we refined our guesses by referring to our error assumptions and considering the maximum accuracy we could expect to achieve. We estimated a typical error on each data spectrum by creating a reference error that was 1\% of the median flux of the spectrum within the wavelength range. Then, we scaled that error value per pixel as $\sigma_{flux} = \text{reference error} * \sqrt{\frac{\text{median flux}}{\text{pixel flux}}}$, where the square root term is the noise term relating the median flux and the individual pixel's flux, assuming the flux is dominated by Poisson noise. Since our observations were synthetic, each pixel was represented as a point on the spectrum, with three pixels assigned per resolution element, as discussed in Section \ref{sec:make spec}. Given these assumptions about the noise, we wished to carefully explore the parameter space down to values that would produce a $<1\%$ variation in the parameters of interest. For normalization, this meant that our small step size standard deviation was 0.5\% of the best fit normalization value, to improve over our 1\% error by a factor of 2. In \teff, to reach an accuracy of 10\% requires a temperature measurement with an accuracy of 100K, so to reach 1\% requires a measurement within 10K. We varied \teff\ by 5K, to improve on the 1\% error by a factor of 2. Finally, for extinction, via a similar argument, to reduce variability to less than 1\%, we needed to determine extinction to $< 1\%$ or 0.01 mag, which we reduced to 0.005 mag to improve on the error by a factor of 2.

\begin{figure*}
\plotone{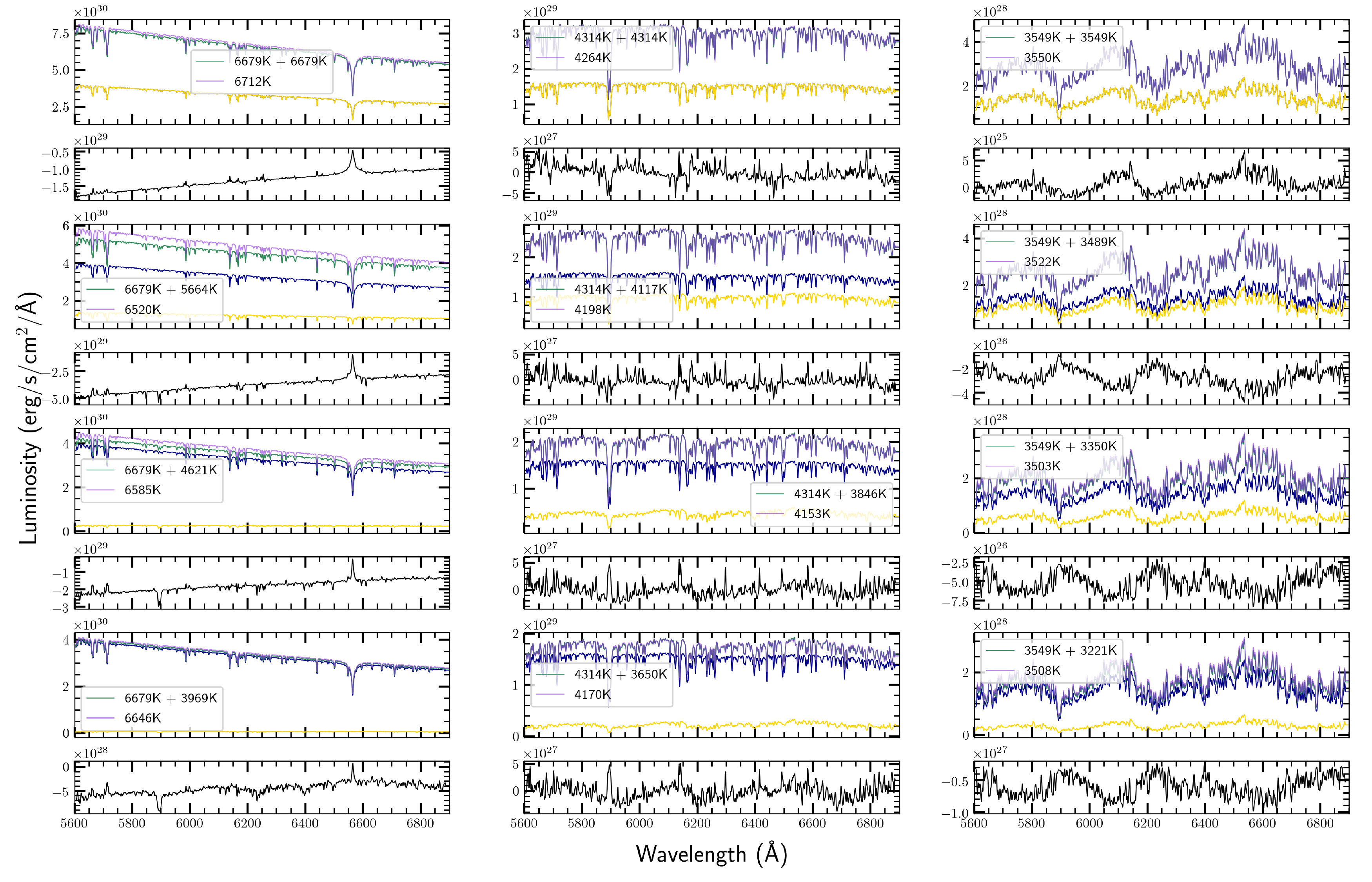}
\caption{The same as Figure \ref{fig:example specs}, with the best fit single star spectrum overlaid on the composite spectrum in purple. The composite spectrum is typically not visible because the match with the best fit spectrum is good enough that it is completely obscured by the fit spectrum. The black lines under each spectrum show the residuals from each fit (data - model). In general there are few spectral features left in the residuals, though the fit is not as good for the F star.}
\label{fig:fits}
\end{figure*}

Figure \ref{fig:fits} shows the same systems as Figure \ref{fig:example specs} with the best fit spectrum overlaid. The fitted spectrum is typically a close enough match to the composite system spectrum that the two spectra are not distinguishable. 

In addition to finding the best fit temperature, normalization, and extinction, we had to determine a system luminosity in order to find a system age and mass. To do this, we took the assigned absolute magnitudes taken from the evolutionary models, converted them to apparent magnitudes using the distance assigned to the system, and then converted them back into absolute magnitudes using an assumed distance for our ``star forming region'', and our measured A$_{V}$ for runs with extinction. We tested two different scenarios: one in which distances are known exactly, by using the assigned system distance as the assumed system distance, and another scenario in which distances are known to $\sim$ 20 pc accuracy. For the less precise distances, we assumed all stars fell at a distance of 140 pc, which is the nominal distance to US. This replicates the lack of precise parallaxes that characterized most literature analyses prior to \textit{Gaia}. 

For each age in the evolutionary models, we interpolated the VRIJHK magnitudes (converted into fluxes) and the luminosity as functions of temperature, then drew a best fit set of magnitudes and luminosity using the best fit temperature, which was measured in the spectral fitting step. At that point we had a luminosity and set of magnitudes for each age at a given temperature. Then, for each filter, we interpolated the luminosity as a function of flux, and drew a best-fit luminosity in that filter using the ``observed'' flux. This resulted in six candidate luminosities produced from the six filters used. We chose the median of the candidate luminosities as our best fit luminosity. After finding a best fit luminosity, we bilinearly interpolated the evolutionary models in temperature-luminosity space to find the system's apparent mass and age. The output system parameters from the fitting process and subsequent inferences are summarized in Table \ref{tab:params}.

\section{Results}\label{sec:results}
We simulated several different populations to test the performance of our simulation code. In our test populations, to minimize the scatter caused by a significant age distribution, the ages were assigned using a Gaussian distribution with $\mu = 10$Myr and $\sigma = 1$ Myr. For the full science simulations, ages were drawn from a Gaussian distribution with  $\mu = 10$Myr and $\sigma = 2$ Myr. 

We tested populations of single stars with no extinction, to assess the quality of our parameter retrievals and establish a baseline of expectations for subsequent simulations. We then simulated a single star population with extinction, to test how well we could measure extinction without the additional errors introduced by the presence of undetected binaries. Finally, we simulated a population of single and undetected binary stars, with extinction, and analyzed it with both exact and inexact distances. We used the exact distance scenario to assess the impact of multiplicity on age measurements without added bias from distance uncertainties. We used the inexact distance scenario, where system distance was assigned using a Gaussian distribution of $\mu = 140$pc and $\sigma = 20$pc, but with the analysis distance held constant at 140 pc for all systems, to understand the impact of the observational uncertainty in pre-\textit{Gaia} distances on the observed binary properties.

\subsection{Single Stars, No A$_{V}$}

\begin{figure*}
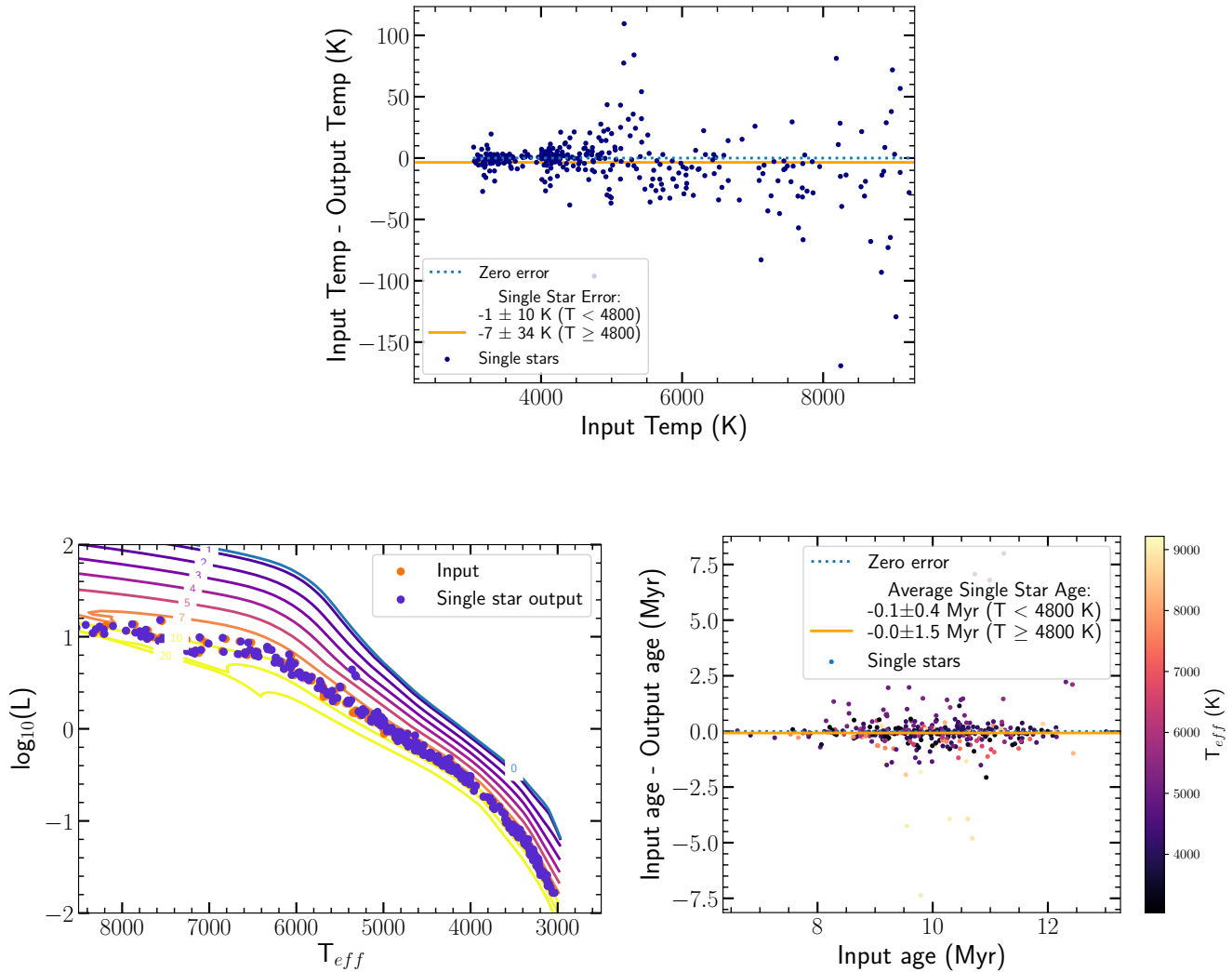

\gridline{
    \fig{singles_temp_plot.pdf}{0.49\textwidth}{}
		}
\gridline{
	\fig{singles_isochrone.pdf}{0.49\textwidth}{}
    \fig{singles_age_diff_plot.pdf}{0.49\textwidth}{}
    }
\caption{Fit summary plots for single stars without extinction. \textbf{Top Left:} Temperature residuals vs. input \teff\ for single stars. The average error is 1 K with an RMS error of 10 K for \teff $<$ 4800 K, and -7 K with an RMS error of 34 K for \teff\ $\geq$ 4800 K. \textbf{Bottom Left: }An HR diagram with MIST isochrones underlaid, showing the input (orange) and output (purple) \teff\ vs luminosity for single stars. The input and output match well. \textbf{Bottom Right: }Age residuals (input - output age) vs. the input age for each single star. The average age error is -0.1 Myr with an RMS of 0.4 Myr for \teff\ $<$ 4800 K, and 0.0 Myr, with an RMS of 1.5 Myr for \teff\ $\geq$ 4800 K.}
\label{fig:single no av}
\end{figure*}

We began by examining a small population of single stars without extinction, to test our accuracy in retrieving the age, \teff, and mass for the stellar population. We expected to be able to recover the input stellar parameters exactly (in the limit of an infinite runtime for fitting) since the population had no uncertainty or binaries. We simulated a population of 360 single stars with a Gaussian age distribution of $\mu = 10$Myr and $\sigma = 1$Myr.

The top panel of Figure \ref{fig:single no av} shows the \teff\ residuals (input \teff\ - output \teff) vs. the input \teff. The average residual value is -1 K with an RMS error of 10 K for \teff\ $< 4800$K, and -7 K with an RMS of 34 K for \teff\ $\geq$ 4800 K. Although the synthetic data have no noise added, and so the fit should be exactly retrieved, the low resolution means that some subtle \teff\ information is blended or otherwise hard to observe, and so the temperatures converge back to the true value slowly. The hotter stars exhibit more scatter because there are fewer temperature-sensitive lines at higher temperatures than at 3000 - 5000 K, where there are TiO bands up to $\sim$ 4300 K and an Na doublet that is sensitive to both temperature and log(g).

A \teff\ - luminosity diagram, with MIST isochrones underlaid, is shown in the bottom-left panel of Figure \ref{fig:single no av}. The input values are shown as orange points, and the output as purple points. The output values fall around the 10 Myr isochrone, as expected, with scatter that is only slightly larger than the 1 Myr age dispersion assigned to the input population.

The bottom-right panel of Figure \ref{fig:single no av} shows the age residual (input-output age) plotted against the input age. We found that the average age error was -0.1 Myr with an RMS of 0.4 Myr for \teff $<$ 4800 K, and 0.0 Myr with an RMS of 1.5 Myr for \teff\ $\geq$ 4800 K, which is less scatter than the typical uncertainties in real observations, even in the less accurate high-temperature regime. After analyzing the simplest possible population, single stars with no extinction, we found that our fitting technique is very accurate, recovering population-wide properties precisely, resulting in average age measurement error of $< 0.1$ Myr, and average temperature error of $< 10$ K. Individual stellar properties are also recovered precisely, with small RMS scatter for both temperatures and ages.

\subsection{Single Stars With A$_{V}$}
\begin{figure*}
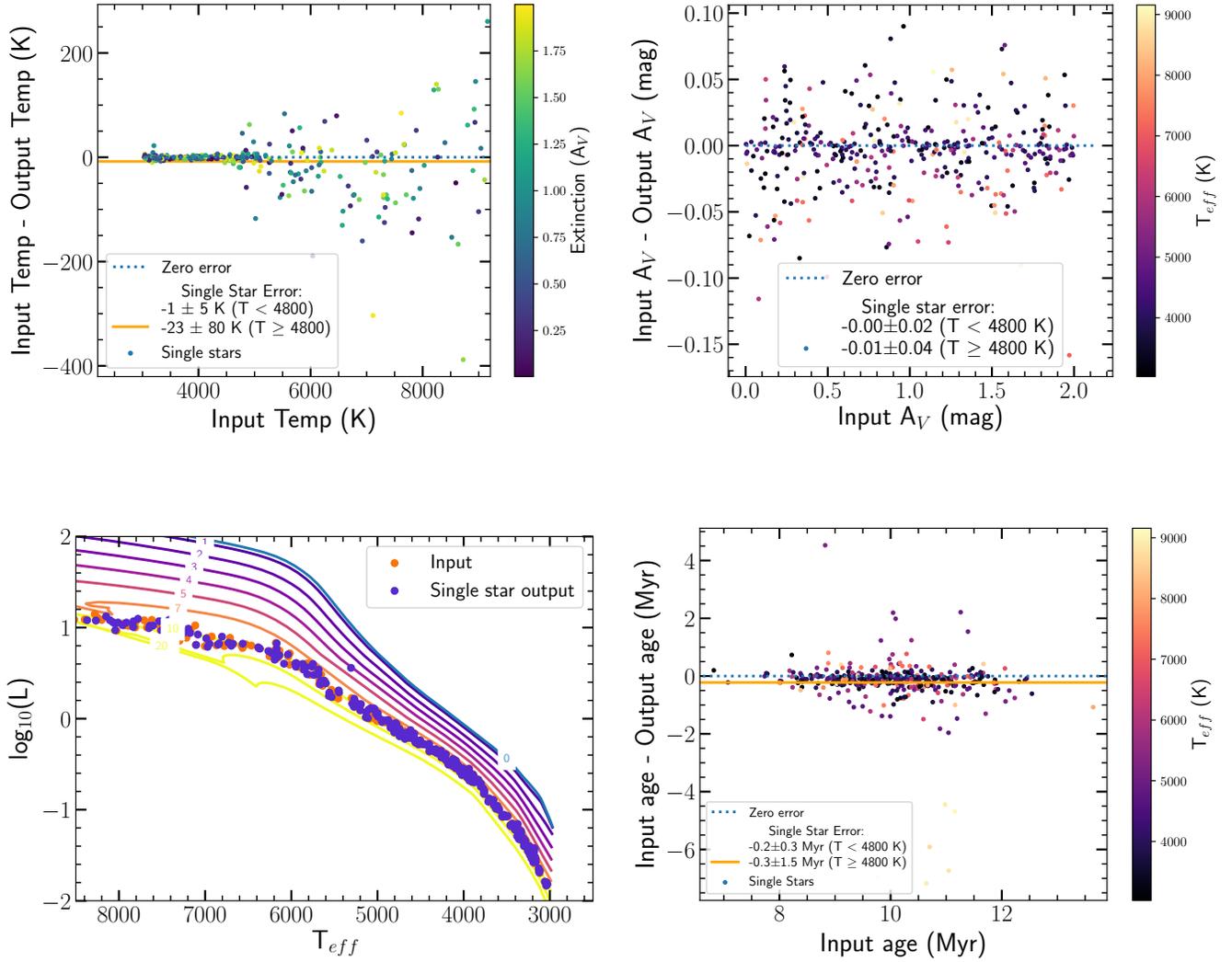

\gridline{\fig{singles_av_temp_plot.pdf}{0.49\textwidth}{}
          \fig{singles_av_av_plot.pdf}{0.49\textwidth}{}
          }
\gridline{\fig{singles_av_isochrone.pdf}{0.49\textwidth}{}
          \fig{singles_av_age_diff_plot.pdf}{0.49\textwidth}{}
          }
\caption{Fit summary plots for single stars with extinction. \textbf{Top left: }Input primary or single star temperature vs. \teff\ residuals between primary star temperature and fitted system temperature.. The average temperature error is -1 K with an RMS error of -5 K for \teff $<$ 4800 K, and -23 K with an RMS error of 80 K for \teff\ $\geq$ 4800 K. The colorbar shows the extinction value for each point. There is no observed correlation between extinction value and temperature residual, indicating that the A$_{V}$ is determined well enough that it does not bias the \teff\ results. \textbf{Top right: }The extinction residual vs. input extinction value. The residual has an average value of 0.00 mag with an RMS of 0.02 mag for \teff $<$ 4800 K, and -0.01 mag with an RMS of 0.04 mag for \teff $\geq$ 4800 K. The colorbar shows the input primary star temperature. \textbf{Bottom Left: }HR diagram with MIST isochrones underlaid. The input stars are orange points, while the output systems are purple. Extinction in the spectra does not significantly affect retrieval of parameters. \textbf{Bottom right: }Age residual vs. input age. The average age error is -0.2 Myr with an RMS of 0.3 Myr for \teff\ $<$ 4800 K, and -0.3 Myr with an RMS of 1.5 Myr for temperature $\geq$ 4800 K. The presence of extinction does not change the observed spread in ages. Hotter stars show more scatter than cooler stars, due to the worse temperature recovery in hotter systems.}
\label{fig:single ext teff}
\end{figure*}

We next imposed variable extinction ($ 0 < A_{V} < 2$) on a sample of 360 single star systems, with a Gaussian age distribution of $\mu = 10$Myr and $\sigma = 1$Myr. We expected to find that large extinction and temperature errors were correlated, and therefore expected hotter stars to have larger extinction residuals. In general, we expected to recover extinction accurately, because our fitting method preserved the shape of the stellar continuum.

The top left panel of Figure \ref{fig:single ext teff} shows the temperature residuals vs. the input temperature for the full population of stars, with each point color coded to reflect its extinction. There is no observed correlation between extinction value and temperature recovery error. The average temperature residual is -1 K with RMS of -5 K for \teff\ $<$ 4800 K, and -23 K with an RMS error of 80 K for \teff\ $\geq$ 4800 K. The temperature residual distribution for stars with extinction is similar to the average error and scatter of the stellar population generated without extinction. 

The top right panel of Figure \ref{fig:single ext teff} shows the extinction residual vs. the input extinction. The extinction residuals have an average value of 0 mag with an RMS of 0.02 mag for \teff\ $<$ 4800 K and -0.01 mag with an RMS of -0.04 mag for \teff\ $\geq$ 4800 K. The extinction residuals have no dependence on input A$_{V}$ or system temperature. Even high temperature systems, which typically have a less accurate temperature recovery, have accurately recovered extinctions, with an RMS scatter that is only slightly larger than the low-temperature population.

The bottom left panel of Figure \ref{fig:single ext teff} shows a temperature-luminosity plot of the input and output populations with MIST isochrones underlaid. This figure again demonstrates that a nonzero extinction does not reduce the quality of our temperature and luminosity recovery, because the points again fall along the 10 Myr isochrone with a small scatter induced by the 1 Myr age dispersion, but no additional scatter.

Finally, the bottom right panel of Figure \ref{fig:single ext teff} shows the age residuals vs. the input age. The average age error was -0.2 Myr with an RMS of -0.3 Myr for \teff\ $<$ 4800 K and -0.3 Myr with an RMS of 1.5 Myr for \teff\ $\geq$ 4800 K. This scatter and mean error is comparable to that in the single star with no extinction case. Hotter stars still show greater scatter than cooler stars, which is due to the less accurate temperature recovery for hotter stars, but the extinction recovery is not correlated with temperature accuracy. We conclude that we can recover the extinction value both accurately and precisely, even in scenarios with a large temperature residual, and the slight differences between assigned and measured A$_{V}$ values do not significantly affect the observed age distribution.

\subsection{Single and Binary Stars, with A$_{V}$}
\begin{figure*}
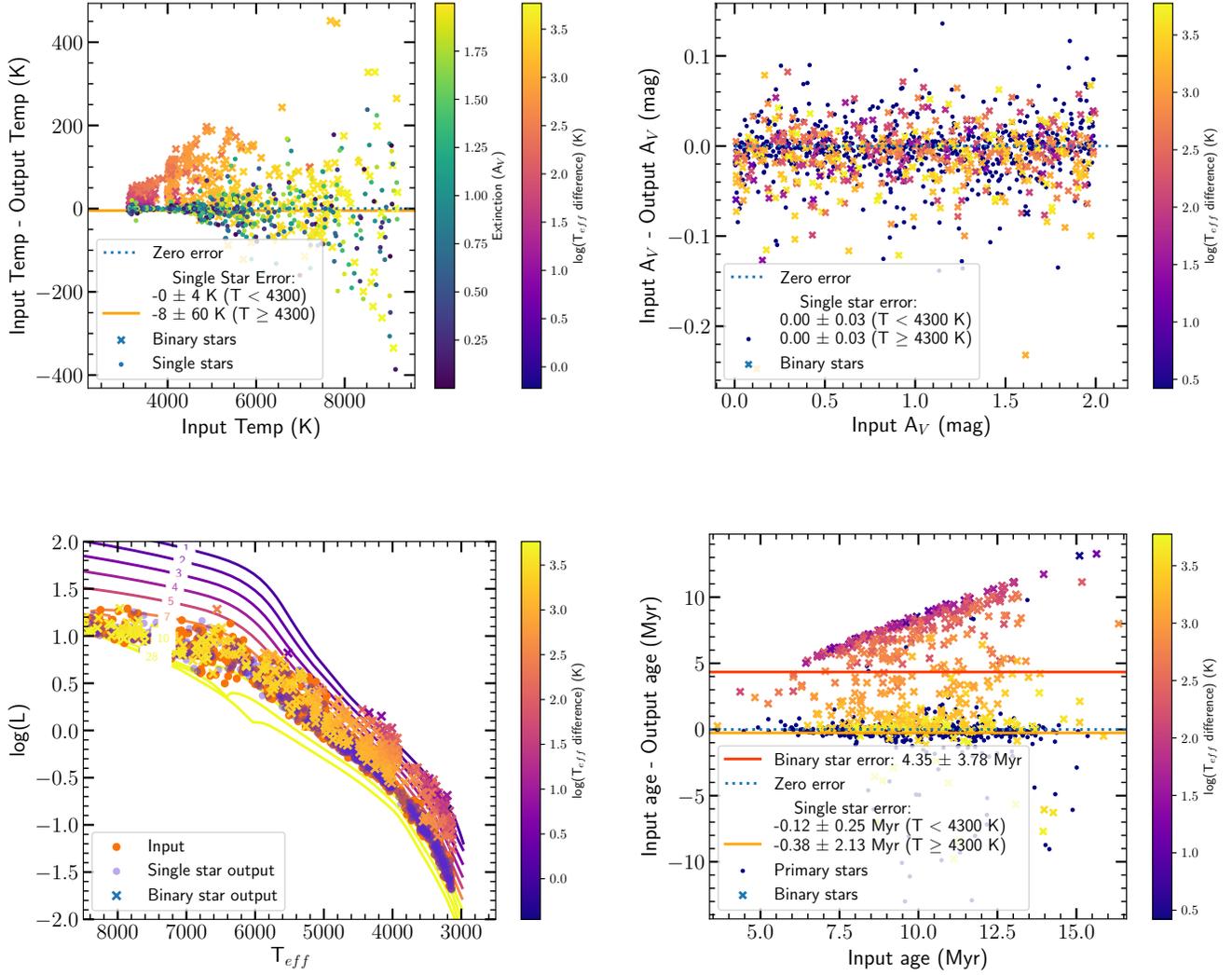

\gridline{\fig{binaries_exact_temp_plot.pdf}{0.49\linewidth}{}
	\fig{binaries_exact_av_plot.pdf}{0.49\linewidth}{}
	}
\gridline{\fig{binaries_exact_isochrone.pdf}{0.49\linewidth}{}
	\fig{binaries_exact_age_diff_plot.pdf}{0.49\linewidth}{}
	}
\caption{\textbf{Fit summary plots for a 1200 system subsample of the exact-distance population of }single and binary stars with extinction. \textbf{Top Left:} Input primary star (crosses) or single star (points) temperature vs. \teff\ residual (input-output temperature). The average temperature error is 0 K with an RMS of 4 K for \teff\ $<$ 4800 K, and -8 K with an RMS of 60 K for \teff $\geq$ 4800 K. The first color bar shows the A$_{V}$ for each single star, while the second color bar indicates the temperature difference between the primary and secondary stars in each binary. Binary stars systematically have a positive temperature residual: the presence of an unknown secondary star biases temperature measurement. \textbf{Top Right: }Input extinction vs. extinction residuals for binaries (crosses) and single stars (points). Multiplicity does not affect the quality of the A$_{V}$ retrieval: we find an average error of 0.00 mag with RMS of 0.03 mag for all temperatures, which is the same as the single star result. \textbf{Bottom Left:} HR diagram of input primary and single stars (purple points) and output single stars (orange points) and unresolved binaries (crosses). The color bar shows the temperature difference between the two binary components. Binaries with a small temperature difference are the most offset from the input, because the flux ratio is large enough for the secondary spectrum to impact the composite spectrum. Low mass binaries appear younger than high mass binaries. \textbf{Bottom Right: }Input age vs. age residuals for both binaries (crosses) and single stars (points). Multiple stars are systematically measured to be younger than single stars, with an upper envelope appearing because the age cannot be negative. Binaries have an average age residual of 4.35 Myr with an RMS of 3.78 Myr, while single stars have an average residual of -0.12 Myr with an RMS of 0.25 Myr for \teff $<$ 4800 K and -0.38 Myr with an RMS of 2.13 Myr for \teff $\geq$ 4800 K, which is the same as the single-star-only case.}
\label{fig:binary_ext_teff}
\end{figure*}

To explore the effect of binary stars on the apparent age distribution of Upper Sco, we simulated a population of 20,000 systems consisting of single and binary stars created using the binary statistics described in Section 2. In the following two subsections we plotted a subset of 1200 systems from the full sample to increase readability of the figures. Observed binary statistics by definition do not include undetected binaries, so may not accurately reflect the true binary fraction. Therefore, our simulation might slightly under-represent the number of undetected binaries. Our goal in this subsection was to measure the offset in the derived stellar parameters, especially age, when the multiplicity of some systems is not known. We assigned each system a random extinction with values of $0 < A_{V} \leq 2$ mag, and drew the ages for the population from a Gaussian distribution centered at $\mu = 10$Myr, with a standard deviation of $\sigma = 2$Myr, to match previously suggested age spreads for Upper Sco.

The top left panel of Figure \ref{fig:binary_ext_teff} shows the temperature residuals vs. the input temperature, with the left color bar showing the extinction values for single stars and the right color bar showing the temperature difference between the two binary components. There is no observed correlation between extinction and the quality of \teff\ recovery for single stars. Binaries have systematically cooler measured temperatures than the input primaries, due to the cooler secondary adding lower temperature spectral features to the summed spectrum. We measure a single-star average temperature error of 0 K with an RMS scatter of 4 K for \teff\ $<$ 4800 K, and -8 K with an RMS scatter of 60 K for \teff $\geq$ 4800 K, identical to the other single-star cases. As the temperature difference between the two binary components increases, the temperature residual increases, until the point at which the primary star is bright enough that the secondary star does not meaningfully contribute to the spectrum. At the high-mass end, most systems are not equal temperature, because the mass ratio distribution is sharply peaked toward lower mass ratios. Additionally, the mass-luminosity relation is steep at large masses, so a slightly less massive secondary star will be much fainter, meaning that it will minimally contribute to the luminosity (and thus minimally bias the derived temperature; Figure \ref{fig:example specs}). At the low-temperature extreme (late-M stars), the effect of multiplicity on the measured temperature is small because most binaries are near-equal-mass systems, meaning that even thought the secondary contributes a significant amount of flux, the temperatures are relatively similar, so the spectrum is relatively unaltered by the presence of the secondary star. At intermediate masses, the secondary star contributes a meaningful amount of flux, and the temperatures are significantly different, so the temperature measurements are biased toward cooler \teff.

The top right panel of Figure \ref{fig:binary_ext_teff} shows the input A$_{V}$ vs. the A$_{V}$ residual for both single stars (dots) and binaries (x symbols). We expected that the addition of a companion would affect the SED by making it redder, so that multiplicity would affect the recovered A$_{V}$, but we found that multiplicity does not impact the quality of the A$_{V}$ fit. Even in the case of a binary system fit as a single star, we can recover extinction accurately, with an average residual of 0.00 mag with RMS of 0.03 mag for all temperatures. The secondary star may contribute enough temperature-dependent features to the spectrum for the fit to attribute the redder SED to a cooler temperature, rather than larger extinction.
 
The bottom two panels of Figure \ref{fig:binary_ext_teff} show two different representations of the bias introduced by binary stars into the age measurement. The bottom-left panel contains a temperature-luminosity diagram with MIST isochrones underlaid, demonstrating that the binary stars systematically fall above the sequence of single input and output stars. The spread is most pronounced at low masses, and binaries with a relatively small temperature difference show the most offset from the input systems: A low-mass small-temperature-difference system is the scenario in which the secondary star flux becomes the most significant to the composite spectrum, and thus has the largest effect on the measured temperature by contributing cooler spectral features. At high temperatures, the binaries are not significantly offset from the single systems. This is because luminosity goes as T$^{4}$, so even a small temperature difference at high temperatures results in a large flux difference. Also, as discussed above, the mass ratio distribution of low-mass stars is peaked toward unity, and the mass ratio distribution of high-mass stars is peaked toward zero (Figure \ref{fig:statplot}). These two factors combine to make the contributions of binaries to age bias at high temperatures negligible.

The bottom right panel of Figure \ref{fig:binary_ext_teff} shows the age residual vs. the input age for single and binary stars. The upper envelope in binary ages is because the points rise above the youngest isochrone in the HR diagram and the age is not allowed to be smaller than the youngest isochrone (an age of 1 Myr). Binaries systematically appear to be significantly younger than the single stars: binaries have an average age residual of 4.35 Myr with scatter of 3.78 Myr, while the single stars still have an average age residual of -0.12 Myr with an RMS of 0.25 Myr for \teff\ $<$ 4800 K, and -0.38 Myr with an RMS of 2.13 Myr for \teff $\geq$ 4800 K, which is identical to the other single star results. The large scatter for the binary measurement occurs because the varied flux ratios of binaries bias the age measurements to different degrees - a small flux ratio will not bias the measured age much, but a relatively large flux ratio means that the secondary contributes meaningfully to both the measured luminosity and temperature, making the age measurement less accurate. We conclude that binary stars have an observational signature that pulls the average age of a population younger, and becomes more pronounced for lower masses, meaning that binarity could explain the mass-dependent trend in measured ages of young stars.

\subsection{Single and Binary Stars with A$_{V}$ and Inexact Distances}

\begin{figure*}
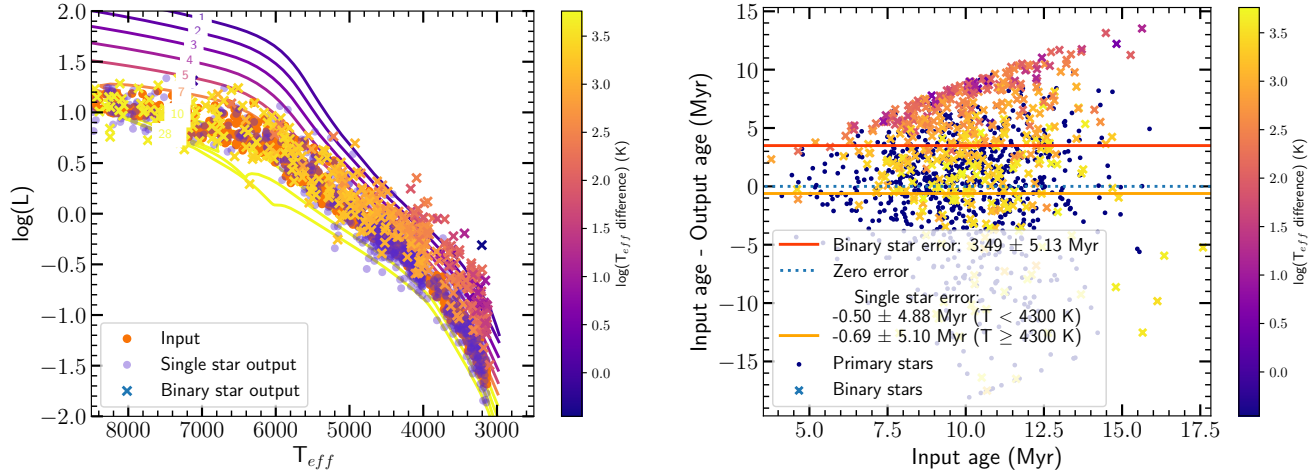

\gridline{\fig{binaries_inexact_isochrone.pdf}{0.49\linewidth}{}
	\fig{binaries_inexact_age_diff_plot.pdf}{0.49\linewidth}{}
	}
\caption{\textbf{Result summary figures for a 1200 system subsample of the inexact-distance population.} \textbf{Left:} The same as the bottom left panel of Figure \ref{fig:binary_ext_teff}, but for the same systems analyzed with inexact distances. The distance uncertainty introduces more scatter into the HR diagram across all masses. \textbf{Right: }The same as the bottom right panel of Figure \ref{fig:binary_ext_teff}, but for an analysis with inexact distances. The added scatter from the distance uncertainty increases the spread in recovered ages for both single and binary stars, blending the two populations together, but the binaries still appear younger than the single stars. We measure an average age residual of 3.49 Myr with an RMS of 5.13 Myr for the binary stars, an age residual of -0.50 Myr with an RMS of 4.88 Myr for the single stars with \teff $<$ 4800 K, and a residual of -0.69 Myr with RMS of 5.10 Myr for single stars with \teff\ $\geq$ 4800 K.}
\label{fig:binary_inexact}
\end{figure*}

In the pre-\textit{Gaia} era, distance uncertainties presented a potential confounding effect, and may have made it more difficult to identify the effects of multiplicity by obscuring the binary sequence in an isochrone, effectively blurring out the effects of binaries on age distributions. We analyzed the same binary population as above: a 1200 system subset of our full 20,000 system sample and a Gaussian age distribution with $\mu = 10$Myr and $\sigma$ = 2 Myr. Instead of analyzing these systems with an exact distance, we assumed that every system had a distance of 140 pc, which is the canonical distance to Upper Sco \citep[e.g.,][]{Rizzuto2015}. Since our systems were generated with a Gaussian distance distribution centered at 140 pc with a standard deviation of 20 pc, this was equivalent to introducing a systematic distance error with a standard deviation of 20 pc. 

We only show the age results for this analysis, because the temperature and extinction fitting steps are not affected by the assumed distance, and the population was exactly the same as that shown in Section 4.3, just analyzed with inexact distances (with no other modifications to the systems). Thus, the temperature and extinction results are identical to those shown in Figure \ref{fig:binary_ext_teff}.

Figure \ref{fig:binary_inexact} shows an HR diagram with isochrones (left panel) and the input age vs. age residual plot (right panel). In both plots, relative to the same figures in Figure \ref{fig:binary_ext_teff}, the trends are the same but with more scatter in both the single and binary star populations, with the ultimate effect of obscuring some of the effects of multiplicity. For example, in the HR diagram shown in the left panel of Figure \ref{fig:binary_inexact}, the binary sequence and age gradient of binaries is still apparent, but the increased spread in both populations means that more single stars fall on the isochrones that were previously only occupied by binaries, and some binaries fall closer to the single-star sequence than they may have without distance uncertainty. This is because in some cases the overluminosity (due to multiplicity) is canceled out by a distance error (e.g., a system is thought to be closer than it truly is), meaning that the binary is pulled down in \teff-luminosity space, onto the single star isochrone. Similarly, a single star with an incorrect distance may appear overluminous and be shifted up in \teff-luminosity space, ending up in parameter space that was previously occupied by binaries. Thus, the two populations become more homogeneous, although the population separation between single and binary stars is still present. 

The right panel of Figure \ref{fig:binary_ext_teff} shows the age residuals vs. the input age. We find that the average binary star age error is 3.49 Myr with scatter of 5.13 Myr, while the single star age error is -0.50 Myr with RMS of 4.88 Myr for \teff\ $<$ 4800 K, and -0.69 Myr with scatter of 5.10 Myr for temperatures $\geq$ 4800 K. The binary star age offset is slightly smaller than in the exact-distance case, and has an age residual spread that is broader by $\sim$ 1 Myr than the exact-distance scenario. The single star population has a similar average age for both the high and low temperature subsets, but the spread in the age residuals is 4.8 Myr and 2.5 Myr larger than the exact-distance scenario for the cool and hot single star populations, respectively. The increased spread is a result of the inexact distance analysis, which causes more spread in all populations (binaries, single stars with temperature $<$ 4800 K, and single stars with temperature $\geq$ 4800 K. Similar to the exact-distance case, we find that binary stars both pull the mean age of the full population younger, and introduce a mass-dependent age gradient, but the binary sequence is less distinct with distance error. As a result, the binaries become less prominent in the HR diagram, meaning that they are more difficult to identify and remove. 

\section{Discussion}\label{sec:disc}

We have found that unresolved binary stars introduce a mass-dependent age gradient into a synthetically observed single-age stellar population because of the mass dependence of the multiplicity fraction and the mass ratio distribution, as well as the shape of the mass-luminosity relation. In this section we begin by discussing the large-scale changes binaries introduce into observed population statistics. We then compare our results to observations of Upper Sco, and then address the impact of our results on future age studies, as well as on the interpretation of past work. We find that our results are consistent with previous observations, indicating that Upper Sco may be a single-age stellar population where past age measurements have been impacted by the presence of undetected binaries. The figures in the previous section showed subsets of the full 20,000 system population to enhance readability of the figures, but the following analysis was performed with the full set of systems to reduce the Poisson noise.

\subsection{Binaries Alter Observed Population Statistics}
Since unresolved binaries are more luminous than a single star of the same spectral type, it is expected that they will affect the observed HR diagram and will be measured to be younger than their single-star counterparts. An unresolved binary may also affect the inferred mass of a system. However, the isomass tracks at young ages are largely vertical (except when FG stars are moving from the Hayashi track to the zero-age Main Sequence), meaning that the effect of unresolved multiples on the measured mass distribution should be small. In this section we discuss the observed age distribution, finding that the age distribution is affected by binaries and that the measured age is spectral-type dependent, producing a statistically-significant decrease in measured age at later spectral types. 

\begin{figure}
\plotone{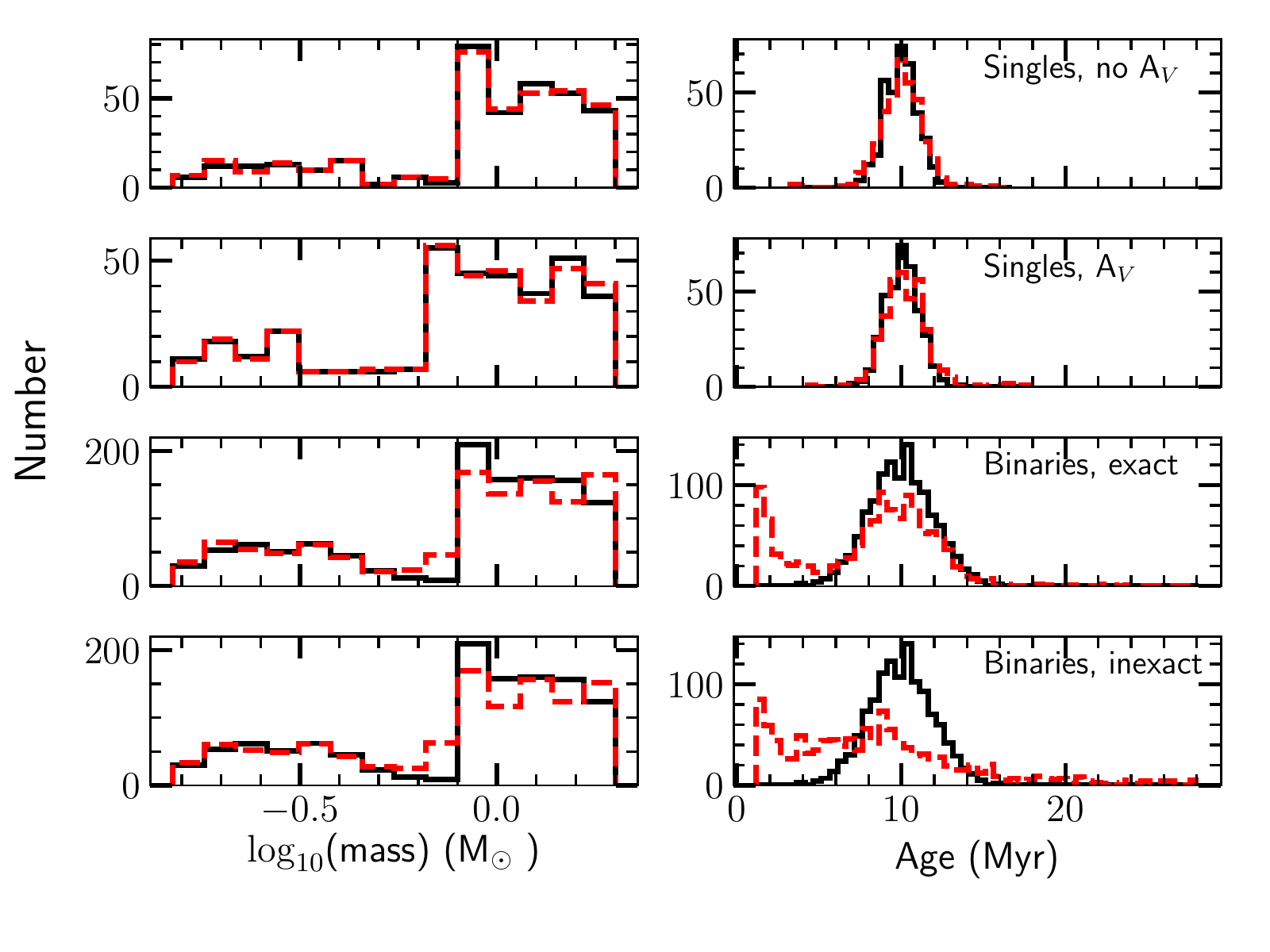}
\caption{Top to bottom: single stars, single stars with extinction, binary stars with exact distance analysis, binary stars with inexact distance analysis. Primary star mass and age histograms for the input systems (black solid line) and output fits (red dotted line). The mass and age distributions change as a result of the presence of undetected binaries: the changes in the mass distribution are small, but the median age is pulled younger and the morphology of the distribution changes.}
\label{fig:outpars}
\end{figure}

Figure \ref{fig:outpars} shows the output mass and age distributions plotted with the input distributions underneath for all of our test populations (single stars, single stars with extinction, and binaries with exact and inexact distances). For the populations of single stars, mass, age, and extinction (where relevant) are all recovered well. In both simulations with binaries, the presence of undetected multiples alters the observed mass and age distributions. There is a slight overabundance of intermediate-mass ($\sim$ 0.9 M$_{\odot}$) stars relative to the input population, and a large population of stars that appear younger than the input population, regardless of distance error. The overabundance of $\sim$ 1 M$_{\odot}$ stars is a result of many different combinations of K/M binaries producing a similar spectrum, while the apparent young population of stars is caused by the undetected binaries. The population with inexact distances shows smoother variation in recovered age than the exact-distance scenario. We observe an output population with a large number of stars that appear $\sim$ 5 Myr younger than the true age of the systems, with many reaching the 1 Myr lower bound on age estimates from the isochrones.

\begin{figure}
\plotone{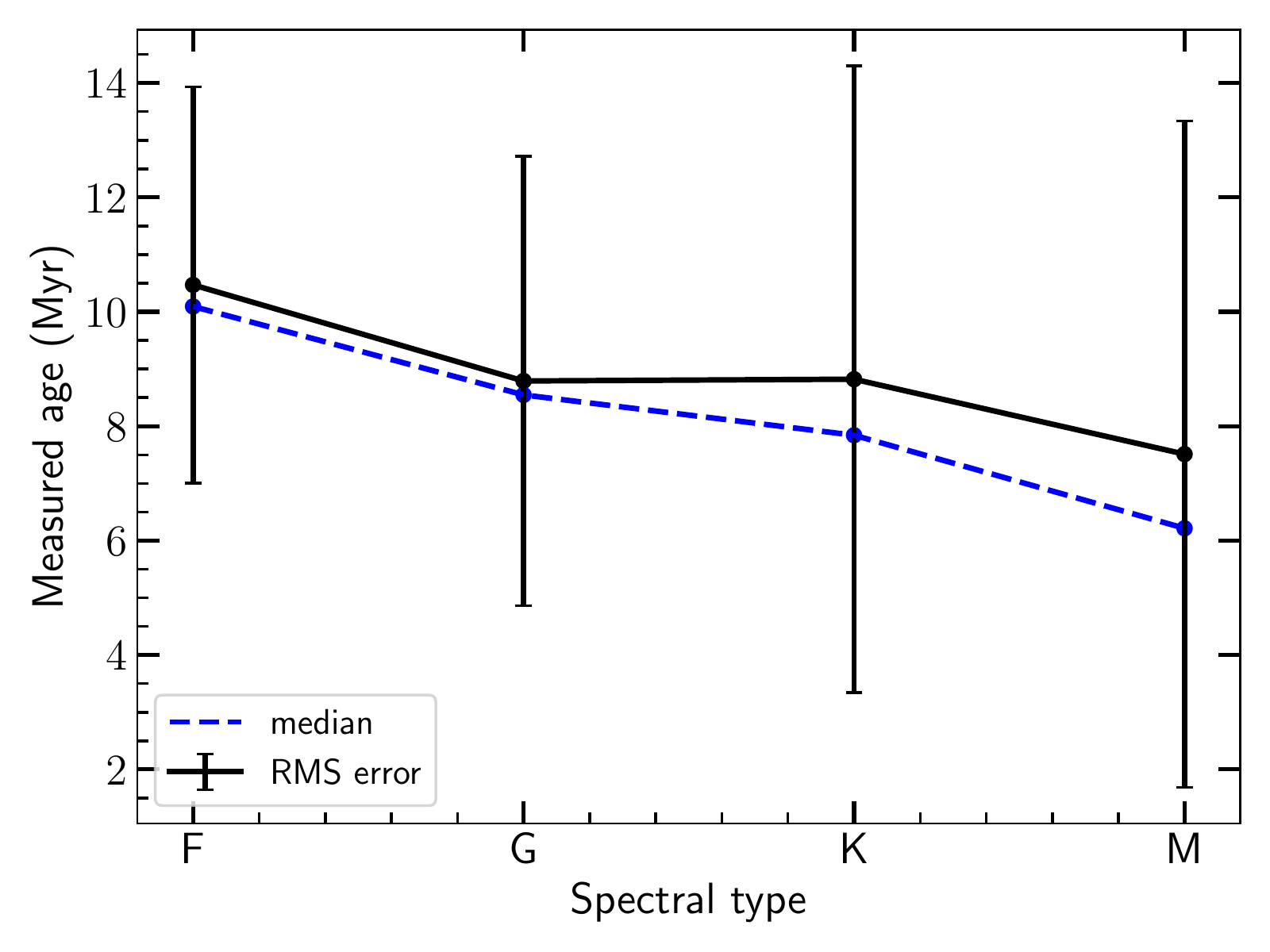}
\caption{Measured median age (blue dotted line), mean age (black line), and RMS error (black error bars) for each measured spectral type in the output population with inexact distances, using temperature to spectral type conversions from \citet{Pecaut2013}. The uncertainty on the mean is smaller than the markers. Later spectral types show more RMS scatter and younger average ages relative to the high-mass stars. The uncertainty on the mean is small, meaning that the measurement of different mean ages for F and M stars is robust, even though the scatter is large.}
\label{fig:age_SpT}
\end{figure}

Figure \ref{fig:age_SpT} shows the mean and median inferred age for each measured system spectral type included in our simulation, using the young-star (5-30 Myr) spectral type-to-\teff\ conversions from \citet{Pecaut2013} to bin the fitted temperatures. The mean age decreases with decreasing mass, and the spread increases, meaning that observations of a population with a large number of binary stars should observe increasing age spreads and younger ages at progressively lower masses. The increasing age spread toward lower masses is due to the progressively larger effect of binaries as mass decreases. We emphasize here that the black error bars do not show an average age error, but instead indicate the observed mean and median age and the standard deviation of the age within a population. The uncertainty in the mean is a function of the sample size, and the simulated sample is large enough that it is negligible.

\begin{figure*}
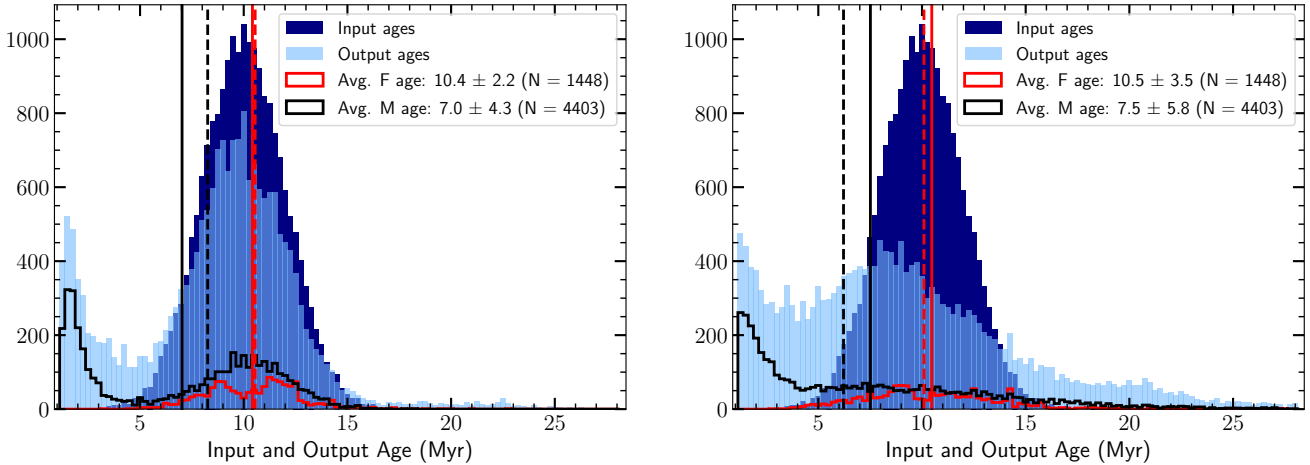

\gridline{\fig{binaries_exact_age_hist.pdf}{0.48\linewidth}{}
	\fig{binaries_inexact_age_hist.pdf}{0.48\linewidth}{}
}
\caption{Input (dark blue) and output (all stars: light blue; M stars: black) age distributions for exact (left) and inexact (right) distance analyses. The mean and median for the F (M) stars are shown as solid and dotted red (black) vertical lines, respectively. In the exact-distance case, the average F star age is 10.4 Myr with RMS of 2.2 Myr, and the average M age is 7.0 Myr with RMS of 4.3 Myr. In the inexact case, the F age is 10.5 Myr with RMS of 3.5 Myr, and the M age is 7.5 Myr with RMS of 5.8 Myr. Binary M stars produce a young peak in both scenarios, biasing the apparent mean age to younger values than for F stars. The bimodality of the M star distribution in the exact distance analysis is somewhat smoothed in the inexact distance analysis, where incorrect distances create more heterogeneity in the measured population ages.}
\label{fig:avg_iso}
\end{figure*}

In Figure \ref{fig:avg_iso}, we show the input and output age distributions for the exact (left) and inexact (right) distance analyses. The input populations are identical because they are based on the same simulated population. The M star output temperature distribution is shown as a black line histogram, and the F star output temperature distribution is shown as a red dotted histogram. The exact distance analysis displays a sharp bimodality in the M star ages caused by the population of binaries. The inexact distance analysis fills in the gap between the two peaks, showing broad scatter with a young peak still produced by the binaries. In both the exact and inexact analysis, some binaries run up against the 1 Myr lower age limit. We suppressed the low-mass stars during system creation, so in reality the age histogram from a survey of the full stellar population of Upper Sco would be dominated by the age distribution of the M stars. However, studies of Upper Sco have typically focused on individual spectral types, where our results are unaffected by the suppression. A full survey of Upper Sco over all spectral types would still observe a bimodality in the age histogram because of the dominance of M stars, which are the population most obviously affected by unresolved multiplicity. 

The F stars recover the correct age for the population. In the exact distance analysis the age spread of the population is accurately recovered as well, with an average F star age of 10.4 $\pm$ 2.2 Myr. In the inexact case the F population has an larger observed age spread of 10.5 $\pm$ 3.5 Myr: A small distance uncertainty can produce a larger age spread in an observed population. The average recovered M star age for the exact-distance population is 7.0 $\pm$ 4.3 Myr, while the average age for the inexact-distance population is 7.5 $\pm$ 5.8 Myr. The measured F and M star average ages are the same for both scenarios, but the inexact-distance population has more scatter in both cases. This is because an inexact distance increases the width of the recovered age distribution for both single and binary stars, while in the exact distance case the scatter is only amplified by the binary stars. Since the binary stars are a sub-population of the total sample, they have less of an effect on the measured scatter. For the remainder of this work we will focus on the inexact-distance population results, because most work in the pre-\textit{Gaia} era had significant distance uncertainties, but we note that the results are also true for the no-distance-error case.

\subsection{Binaries Produce Apparent Age Gradients}
To further explore the cause(s) of the observed age gradient in Upper Sco and other star-forming regions, we compare our age distributions and their properties to those measured by previous observational studies. We find that our results are consistent with past observations of age distributions in Upper Sco, indicating that undetected binaries contribute to observed age gradients in star forming regions. We briefly discuss the results of \citet{Preibisch2002} over the full mass range considered in this study (0.1 $<$ M$_{\odot}$ $<$ 2), then consider the F and G star results from \citet{Pecaut2012}, the M star results of \citet{Rizzuto2015}, and finally briefly discuss the K star ages measured in \citet{Pecaut2016}.

The first large survey for low-mass stars in Upper Sco was conducted by \citet{Preibisch2002} (see also \citealt{Preibisch1999} and \citet{Preibisch2001}). Their survey spanned stellar masses from $0.1 \lesssim M_{\odot} \lesssim 2$, and \citet{Preibisch2002} used an HR diagram to measure an age of 5 Myr with minimal spread after a completeness correction that included factors for unresolved binaries. This result agreed with other, older work, such as \citet{DeGeus1989} and \citet{deZeeuw1985}. 

However, \citet{Pecaut2012} found a median age of 13 Myr with an intrinsic age spread of 3-5 Myr using F stars and 9 Myr with an intrinsic spread of 3-4 Myr for G stars, and adopt an 11 Myr age for Upper Sco. This older age for higher mass stars makes sense in the context of our simulations, as we found that F stars are not significantly affected by the presence of binaries, with average ages and age spreads consistent with the input population (in the zero error case) or slightly broader spreads than the input population (in the distance error case). This is because at higher masses, most binary companions are substantially fainter than the primary stars due to a bottom-heavy mass ratio distribution and the steep temperature-luminosity relation. 

All of the stellar populations considered by \citet{Pecaut2012} showed a slight mass-dependent age trend, and progressively lower-mass stars show younger ages, a trend that is consistent with the effect of multiplicity. The effect of multiplicity should be smallest for the F stars, suggesting that a possible ``true'' age for Upper Sco may be closer to 9-13 Myr with an intrinsic age spread that is likely smaller than the measurement in \citet{Pecaut2012}, after accounting for binaries. The F and G star samples of \citet{Pecaut2012} were small ($\sim$20 F stars, and 16 G stars), so the spread of the observed distribution may be artificially large due to the small-sample sensitivity to outliers, and distance errors introduce additional observed age scatter.

In previous studies of the low-mass stars of Upper Sco, such as \citet{Rizzuto2015}, the M star population showed large scatter, with an observed population of stars falling above the 1 Myr isochrone (i.e., stars that appear younger than 1 Myr). \citet{Rizzuto2015} did not show the age distribution of their M star sample, but the HR diagram of their results shows progressively more scatter and younger ages for lower masses, with late M stars often falling at ages $< 1$ Myr, and most stars appearing younger than 10 Myr. The M stars from our simulation are shown as a black histogram in the age distributions plotted in Figure \ref{fig:avg_iso}, showing a bimodal distribution with the younger peak (age $< 3$ Myr) mostly comprised of M star binaries. The age distribution for the M star population as a whole is broad (spread of $\pm$4-5 Myr), a result of the large impact and high frequency of similar-mass binaries at low masses. 

The scatter and age measurement observed in late M stars in \citet{Rizzuto2015} is consistent with the observational signature of binaries found in this work, both in the age gradient and the observed scatter ($\pm$ 5 Myr in both cases). Distance uncertainty and other observational errors mean that the two distinct populations of M stars (binaries and single stars) are blurred together, creating a unimodal age distribution with a large age spread. Thus, the scatter in M stars observed by \citet{Rizzuto2015} is consistent with the observational signature expected from binary stars, although they also note that their sample is less complete for late M stars and that potentially only the youngest and nearest stars were detected. However, that sample should also have an overabundance of binary stars due to their relative overluminosity, increasing the effect of binaries on the observed population.

Finally, before their corrections for binaries, the K-type sample of \citet{Pecaut2016} had a median age of 5 Myr with scatter (RMS about the mean) of 2 Myr, while we find an average age for K stars of 7.5 Myr with scatter of 3 Myr. We note here that the low-mass stellar suppression in our simulation (intended to improve Poisson statistics for FGK stars) occurs at a spectral type of K6, so the frequency of the lowest-mass K type stars in our simulation are affected by the suppression. The two ages are marginally consistent with each other, and do demonstrate similar age spreads in their ``raw'' populations, which are unaffected by their corrections of observational uncertainty. After the \citet{Pecaut2016} uncertainty corrections, which include corrections for completeness and multiplicity, they suggest that the K star population has an intrinsic age spread (star formation duration) that is Gaussian, with an RMS of $\sigma = $ 7 Myr. This is consistent with the age spread observed in our total simulated population, but is larger than our observed K star age spread. \citet{Pecaut2016} ultimately discard their K star age and use other populations for their age analysis, arguing that their K star results are biased by star spots and other observational effects. Our K star results are marginally consistent with the findings of \citet{Pecaut2016}, but because K stars are the location where our suppression of the low-mass IMF begins, and \citet{Pecaut2016} suggest potential caveats in their K star results, we emphasize that these results are the least reliable out of the various studies we consider.

We conclude that both the population statistics and age distributions for our simulated population are mostly consistent with previous observations of Upper Sco. Thus, our results show that the observed Upper Sco age distribution can be produced without a young stellar population or a true age gradient, support the small age spread previously suggested for Upper Sco, and suggest that unresolved binaries may contribute to mass-dependent age gradients observed in star-forming regions. 

\subsection{Prospects for Future Demographic Studies}
We have found that our results are consistent with observations of Upper Sco, suggesting that the age gradient observed in other star-forming regions may also be caused by unresolved binary stars. In this subsection we discuss other possible sources of the gradient, then consider possible implications of these results on future work.

We found that the observational signature of a population of unresolved binaries is a young peak in the age distribution of a population that as a whole appears to have a large intrinsic age spread. For example, the ages of individual stars in our 10 Myr population fall mostly between the binary peak at 2.5 Myr and the peak of the input distribution at 10 Myr, but with some systems appearing as old as 15-20 Myr and as young as 1 Myr. This scatter suggests that the binary sequence that is often referred to when making HR diagram cuts only represents the most luminous binary systems. Removing those systems simply removes obvious binaries while retaining those that are more difficult to detect, complicating the analysis process. The large scatter in the simulated M star age distribution is consistent with observations of Upper Sco that find an age scatter as large as $\pm 7$ Myr in the low-mass population \citep{Pecaut2016, Rizzuto2015}. 

In the future, better identification of binaries (e.g., using \textit{Gaia} astrometric solutions) will allow analysis and/or removal of binary systems from samples used to measure stellar ages. Shorter period binaries can be identified from astrometric motion \citep[e.g.,][]{Belokurov2020}, while longer-period (optically-resolved) systems can be identified from photocenter jitter and PSF mismatch (e.g., \citealt{Rizzuto2018}; Kraus et al. in prep), or can be resolved directly in the \textit{Gaia} catalog \citep{Ziegler2018, Brandeker2019}. In the interim this work presents a more accurate treatment of binaries that can inform error analyses and interpretation of ages of star-forming regions. 

In addition to distance uncertainty, there are other observational uncertainties, as well as other stellar physics occurring alongside multiplicity. Previous work has suggested that star spots \citep{Rizzuto2015, Pecaut2016} may play a role in observed age gradients, and thus that some physics, such as an appropriate treatment of stellar magnetic fields, may be missing from the models used to infer age \citep{Feiden2016, David2019}. A limitation of the current work is that it does not include a full consideration of the effects of magnetic fields (e.g., star spots), instead modeling each star in a system as a single-temperature object. Future work may address this by including the effects of star spots on a spectrum, in addition to undetected binary stars. Evolutionary models that include magnetic fields, such as the Dartmouth models \citep{Dotter2008, Feiden2012, Feiden2015, Feiden2016} are important steps toward addressing this uncertainty. More studies of spotted stars will also increase our understanding of the appropriate treatment of spots in spectral modeling, making more accurate temperature retrievals for low mass stars possible and thus decreasing the scatter in HR diagrams of star forming regions. A decrease in the scatter of a cluster's HR diagram will make the observational signature of unresolved binaries more evident. However, star spots and unresolved binaries likely need to be addressed simultaneously, because the signatures of star spots and binaries are similar (i.e., both a spot and a secondary star introduce an additional spectral component with a cooler \teff), meaning that this problem is likely quite complicated.

In the \textit{Gaia} era, many observational uncertainties have been significantly reduced. The distance uncertainties that plagued older studies have almost vanished, although the \textit{Gaia} astrometric solutions for unresolved binary stars are less precise, a result of orbital motion and photocenter jitter (Krolikowski et al. in prep). However, with the decreased scatter caused by accurate distances to single stars, HR diagrams of clusters should have clearer binary sequences \citep[e.g.,][]{Luhman2020}. In \citet{Luhman2020} that effect was leveraged to remove obvious binaries, but in the future \textit{Gaia} may instead enable more robust binary statistics for star-forming regions to be performed, both by increasing the sample of close binary stars and facilitating better assessment of cluster membership. This work has emphasized the need for identification and removal of individual binaries, illustrating that a clear binary sequence does not exist. Instead, the apparent binary sequence simply represents a boundary condition for the binary population, with multiples occurring throughout the observed spread of a population in temperature-luminosity space. Removal of an apparent binary sequence simply excises the obvious binaries, while retaining the less-obvious (and thus harder to correct for) binaries. 

A more robust sample of binary stars in clusters will enable better binary statistics for individual star-forming regions. Until the observational scatter in the HR diagram of a cluster can be significantly reduced via smaller observational uncertainties and better evolutionary models and synthetic spectra, future work that measures the age of a star-forming regions should incorporate a careful treatment of binaries. That treatment may include a more complex error analysis that accounts for higher-level binary effects (e.g., by implementing mass-dependent binary statistics), or may simply motivate a more detailed consideration of binaries when interpreting HR diagrams. Regardless of methodology, binaries in star-forming regions must be considered more deeply if better than $\sim$ 5-10 Myr accuracy in an age is desired.

\section{Conclusions}\label{sec:conclusion}
To explore the effects of unresolved binary stars on age measurements in the Upper Scorpius star-forming region, we have performed a low-resolution optical spectroscopic synthetic survey of a simulated star-forming region with a mean age of 10 Myr and a standard deviation of 2 Myr. Our simulated population included unresolved binaries, realistic extinction values, and distance uncertainty, to introduce observational error into our results. We found that unresolved binaries introduce a mass-dependent age gradient and a mass-dependent observed age spread, with lower-mass stellar populations progressively appearing younger with a larger age spread. This is a result of the steep mass-luminosity relation at high masses, the preference for small-mass-ratio binaries at high masses, and the inverse preference for equal-mass-ratio binaries at low masses.

The observed age gradient and age spread are consistent with previous observations of Upper Sco, meaning that our results support the hypothesis of a $\sim$10 Myr age for US, with a small intrinsic age spread. Star spots and other observational effects like model mismatches with data introduce confounding factors and additional errors to age measurements, meaning that the bimodal low-mass age distribution that is the distinct observational signature of a binary population is likely not observable. Instead, undetected binaries will have a similar observational signature as that expected from star spots, creating a mass-dependent age gradient and artificially large age spread. Future work should leverage new observational and modeling resources to reduce observational uncertainties, and in the meantime should more carefully account for binaries in error analyses and interpretation of HR diagrams.

\acknowledgements
We thank the referee for their helpful comments and suggestions. We thank Greg Herczeg, Andrew Mann, Stella Offner, Aaron Rizzuto, and Ben Tofflemire for useful discussions and feedback on the methods and results of this work. K.S. acknowledges that this material is based upon work supported by the National Science Foundation Graduate Research Fellowship under Grant No. DGE-1610403. The authors acknowledge the Texas Advanced Computing Center (TACC) at The University of Texas at Austin for providing high-performance computing resources that have contributed to the research results reported within this paper.

\software{NumPy \citep{Harris2020}, 
Matplotlib \citep{Hunter2007}, 
SciPy \citep{Virtanen2020}}

\appendix
\section{Comparing Temperature and Extinction Measurement Techniques}\label{sec:appendixA}
Our analysis mimics the spectral fitting technique used by \citet{Rizzuto2015}, which simultaneously fits for \teff\ and extinction using a spectrum that is not necessarily flux calibrated, but that has not been continuum normalized. We use the term ``continuum normalization'' to describe the process of fitting a spectrum with a polynomial or spline, then dividing by that fit to produce a normalized, flat spectrum where the average value is 1, while in the body of the text ``normalization'' is the multiplicative factor used to adjust a model spectrum such that it has flux levels that match a ``flux-calibrated'' synthetic spectrum. Normalization may also describe the division of a spectrum by its maximum value such that the maximum value of the spectrum is 1, but in that case the shape of the continuum is preserved.

Rather than simultaneously fitting for temperature and extinction, a common alternative method for measuring A$_{V}$ and spectral types is to first measure the temperature by comparing a continuum-normalized spectrum with a grid of spectral standard stars (which may either be models or empirical templates), then calculating a color excess by comparing the expected colors (from models or empirical colors) to observed photometry. That color excess can then be converted to an A$_{V}$ \citep[e.g.,][]{Cardelli1989, Fiorucci2003}. \citet{Preibisch2002}, \citet{Pecaut2012}, and \citet{Pecaut2016} all use variations of this method to analyze their samples. It can be difficult to flat-field spectra, so a data reduction method that preserves the shape of the continuum is not always possible. Thus, we wished to understand whether the two methods were comparable.

To explore whether the simultaneous and asynchronous methods had the same results, we simulated a small number of single stars with temperatures of 3100, 3700, 4500, 5500, and 6500 K, which corresponds to spectral types of approximately M4, M0, K3, G5, and F4, converting from \teff\ to spectral type using the conversion for young (5-30 Myr) stars in \citet{Pecaut2013}. We extincted each star with 0.1, 0.5, 1, and 2 mag in A$_{V}$, to match the range of extinctions observed in Upper Sco, and so ultimately analyzed 20 example systems (each of the five temperatures with all four values of extinction).

To perform the simultaneous fits of temperature and extinction, we fitted each spectrum using our usual fitting technique as described above in Section \ref{sec:fitting}. We typically found temperature errors of $\Delta T < 20$K, and extinction errors of $\Delta A_{V} < 0.05$ mag. We found slightly larger systematic temperature errors for the hotter stars, which is consistent with our results from large stellar populations (Section \ref{sec:results}), but no systematic trends in temperature or A$_{V}$ errors. 

To perform the asynchronous fits, we first continuum normalized each extincted input spectrum by dividing by a polynomial. We then fit each spectrum with another continuum-normalized template using the method described in Section 3, but with our only fit parameters being temperature and normalization. We generally recovered temperatures to within 10 K, with a slight decrease in accuracy for the hottest (6500 K) spectrum. This is expected, given that we observe systematically worse temperature recovery for hot stars in our large populations (Section \ref{sec:results}).

After finding a best-fit temperature with spectral fitting, we read in synthetic BVJHK$_{S}$ photometry from the MIST isochrones at an age of 10 Myr for the best-fit temperature, and calculated the B-V, V-J, V-H, and V-K$_{S}$ colors. We chose these colors to match the technique of \citet{Pecaut2016}, who used a method with fewer colors than \citet{Pecaut2012} but more colors than \citet{Preibisch2002}. 

We extincted the synthetic intrinsic colors for the best-fit temperature using the appropriate A$_{V}$ and the A$_{\lambda}$-to-color excess conversion of \citet{Pecaut2016} and \citet{Fiorucci2003}. We computed comparison zero-extinction intrinsic colors, corresponding to the best-fit \teff\ value, using the intrinsic colors given for 5-30 Myr stars in \citet{Pecaut2013}. We compared our model extincted colors to the intrinsic colors to compute an ``observed'' color excess, which we then converted back into an A$_{V}$ using the \citet{Fiorucci2003} relation between color excess and A$_{V}$. Mimicking \citet{Pecaut2016}, we took the median A$_{V}$ as the ``true'' value and the standard deviation of the four values (derived from the four colors) as the error on that measurement. For 3100, 3500, 4500, 5500, and 6500 K, we found measurement errors of 0.42, 0.59, 0.44, 0.14, and 0.12 mag, respectively, but each median extinction matched the correct value fairly well.

\begin{figure}
\plotone{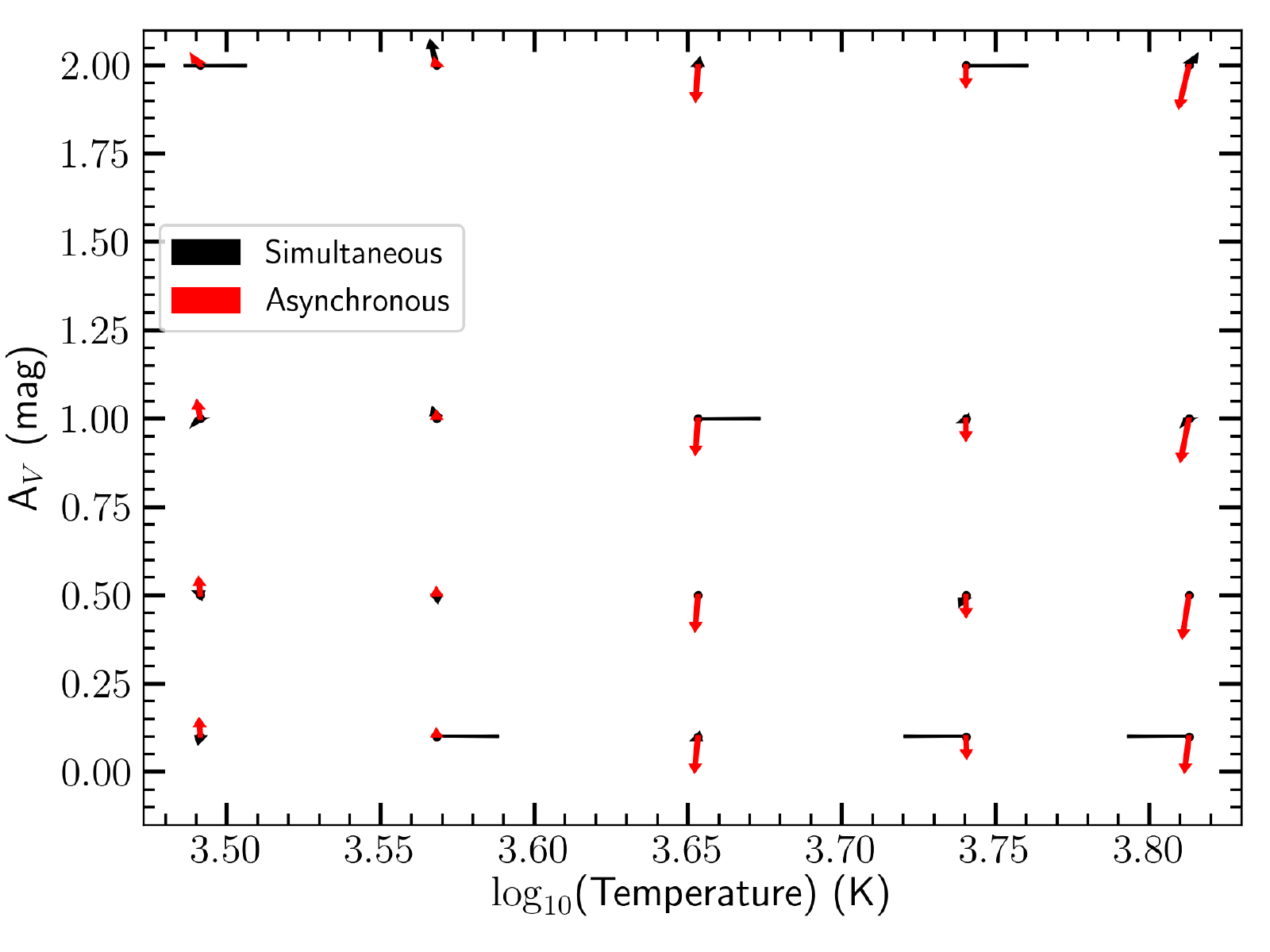}
\caption{Temperature vs. extinction for all 20 test systems for synchronous (black) and asynchronous (red) fits. The starting point of each arrow is the true value, and the end point is the fitted value. The asynchronous fits show systematic trends in extinction and temperature recovery, while the errors for the synchronous fits are typically smaller and more randomly oriented. In both cases the errors are small.}
\label{fig:compare}
\end{figure}

Figure \ref{fig:compare} plots each input (temperature, A$_{V}$) value, with arrows pointing to the resultant fit values for each combination for the synchronous and asynchronous fitting techniques. The errors are typically smaller in both \teff\ and A$_{V}$, but in general all errors are small ($< 0.1$ mag in A$_{V}$, $< 20$ K in \teff). The asynchronous technique shows some systematic trends where results are typically slightly cooler and slightly less extincted than the true values, but generally does not do much worse than the synchronous fitting method.

From Figure \ref{fig:compare} we conclude that the asynchronous and synchronous fitting techniques perform similarly, and so the choice of fitting method will not greatly impact our results beyond neglecting measurement error (which is intrinsic to measuring A$_{V}$ using several different colors). There is a large deviation between different color excess measurements (a large measurement error), but good agreement of the median with the true value, indicating that although single-color extinction measurements may be up to 0.5 mag away from the true value, multi-color measurements are generally reliable. The derived values from the asynchronous method are still in good agreement with the true values, and show little variation between systems with different effective temperatures (the error on the residual is small). Thus, we use the synchronous fitting technique without any loss of our ability to compare our results to those of \citet{Preibisch2002}, \citet{Pecaut2012}, and \citet{Pecaut2016}.

\bibliography{bib.bib}

\end{document}